\documentclass[10pt,a4article]{article}

\usepackage{jheppub}
\usepackage[T1]{fontenc} 
\usepackage{epsfig,epsf}
\usepackage{amsmath}
\usepackage{amsthm}
\usepackage{amsfonts}
\usepackage{amssymb}
\usepackage{caption}
\usepackage{dsfont}
\usepackage{mathrsfs}
\usepackage{epstopdf}
\usepackage{multirow}

\usepackage{rotating}

\usepackage{marvosym}

\usepackage{soul}
\usepackage{xcolor}

\usepackage{caption}
\usepackage{subcaption}
\usepackage{array}   
\newcolumntype{C}{>{$}c<{$}}

\usepackage{slashed}

\usepackage[active]{srcltx}

\def\II{\hbox{{1}\kern-.25em\hbox{l}}}
\DeclareMathOperator{\sign}{sign}
\DeclareMathOperator{\Li}{Li}

\def\II{\hbox{{1}\kern-.25em\hbox{l}}}

\title{
{\textsc
Three-loop evolution kernel for transversity operator}
}

\author[]{A. N. Manashov,}
\author[]{ S. Moch}
\author[]{and L. A. Shumilov}

\affiliation[]{
   II. Institut f\"ur Theoretische Physik, Universit\"at Hamburg,
   D-22761 Hamburg, Germany}

\emailAdd{alexander.manashov@desy.de}

\emailAdd{sven-olaf.moch@desy.de}

\emailAdd{leonid.shumilov@desy.de}

\abstract{
We calculate  quantum corrections to the symmetry generators for the transversity operators in quantum chromodynamics (QCD) in the two-loop approximation.
Using this result, we obtain the evolution kernel for the corresponding operators at three loops.
The explicit expression for the anomalous dimension matrix in the Gegenbauer basis is given for the first few operators.
       }

\keywords{QCD, radiative corrections, re\-nor\-mali\-za\-tion, parton distribution functions}
\preprint{DESY-24-103}

\begin{document}
\maketitle

\newpage
\section{Introduction}
The modern description of hard scattering processes in quantum chromodynamics (QCD) is based on the {\it factorization} approach~\cite{Collins:1989gx} which allows
one to separate short- and long-distance  phenomena. The scattering amplitude of such a process is  given by the convolution of a
coefficient function (hard part) with a non-perturbative quantity (soft part) which can be expressed as the matrix element of  a certain
operator. The scale dependence of the latter is determined by the
renormalization group equation (evolution equation).
The present state of affairs is different for processes with zero and nonzero
momentum transfer between  the initial and final hadron states.
In deep-inelastic scattering (DIS) processes (forward kinematics) the evolution kernels (splitting functions) are known at the
next-to-next-to-leading order (N$^2$LO)~\cite{Moch:2004pa,Vogt:2004mw}
and there are partial results at the N$^3$LO (see~\cite{Cooper-Sarkar:2024crx} and references therein).
The Mellin moments of the splitting functions give the forward-anomalous dimensions~---~the diagonal elements of the anomalous dimension matrix  which enters the renormalization group equation (RGE) for the corresponding local operators.
In processes with a nonzero momentum transfer one has to take into account mixing with total derivative operators which is governed  by an off-diagonal part of the anomalous dimension matrix (off-diagonal evolution kernel).
Calculating evolution kernels directly in off-forward kinematics at high orders demands substantial computational effort and is currently not practical beyond two loops.

An alternative to the direct calculation approach was developed by Dieter M\"uller  in~\cite{Mueller:1991gd,Mueller:1993hg}.
He has shown that the evolution kernel at $\ell$-loops is completely determined by the forward anomalous dimensions and a special quantity, dubbed as a
conformal anomaly, at one order less, i.e. $(\ell-1)$-loops.
Soon after, all evolution kernels of the twist-two operators in QCD were calculated with two-loop
accuracy, \cite{Belitsky:1997rh,Belitsky:1998gc}. A recent development of this method is based on the idea of considering QCD in
non-integer dimensions at a critical value of  the strong
coupling~\cite{Braun:2013tva,Braun:2014vba,Braun:2016qlg,Braun:2017cih,Strohmaier:2018tjo}
to restore the {\it exact} conformal invariance of the theory.
The restoration of symmetry significantly simplifies the analysis, enabling the determination of the evolution kernels of the twist-two vector and
axial-vector operators with three-loop accuracy~\cite{Braun:2017cih,Braun:2021tzi}.

The aim of the  present work is to calculate the evolution kernels for the transversity operators with three-loop accuracy. The nucleon
matrix elements of these operators define the chiral-odd GPDs, see e.g.~\cite{Diehl:2003ny,Belitsky:2005qn}. In deeply-virtual Compton
scattering (DVCS) processes, transversity operators contribute only to the power suppressed helicity-flip amplitudes, making quark-helicity
flip subprocesses  strongly suppressed and chiral-odd GPDs difficult to access experimentally. Nevertheless, their experimental
determination  seems to be feasible in photo- or electroproduction or deeply-virtual meson production processes at energies of the
Electron-Ion Collider (EIC), see e.g. \cite{Beiyad:2010qg,Hyde:2011ke,Cosyn:2021llh,Cosyn:2019eeg,Boussarie:2016aoq}.

Until now the evolution kernel for transversity operators was known with two-loop accuracy. The one-loop  kernel was
derived in~\cite{Belitsky:1997rh}. The two loop expression was obtained
in~\cite{Belitsky:1998gc,Belitsky:1999gu} using conformal anomaly technique. This result was  later confirmed
%
%
by the direct calculation of the two-loop kernel~\cite{Mikhailov:2008my}. Another result for the leading contributions to the anomalous dimension matrix in the limit of a large number of flavors $n_f$ have been
obtained in~\cite{VanThurenhout:2022nmx} at all orders. The forward anomalous dimensions for the transversity operators are known with
three-loop accuracy~\cite{Artru:1989zv,Koike:1994st,Kumano:1997qp,Hayashigaki:1997dn,Vogelsang:1997ak,Velizhanin:2012nm,
Bagaev:2012bw,Blumlein:2021enk}. In what follows we calculate the two-loop conformal anomaly and reconstruct the three-loop evolution
kernel for the transversity operators.

\vskip 5mm

The paper is organised as follows: Section~\ref{S-def} is introductory, we set definitions and notations and give a brief description of
the method used to calculate the evolution kernel. In Sect.~\ref{sect:kernels_and_anomaly} we present the results of calculation  of the
evolution kernel and the conformal anomaly with two-loop accuracy. In Sect.~\ref{s-three-loop} we reconstruct the evolution kernel at
the three-loop level. Explicit expression for the anomalous dimension matrix
in the Gegenbauer basis is given in Sect.~\ref{sect:local}.
Section~\ref{sect:summary} is reserved for summary and outlook. The paper  contains several appendices where  the
analytic expressions for the kernels are collected.

\section{Background} \label{S-def}

Since we are interested  only in the evolution equation it is convenient to work in Euclidean space. The QCD Lagrangian in $d =
4-2\epsilon$ dimension Euclidean space reads
\begin{align}\label{L-QCD}
L=\bar q\slashed{D}q+\frac14 F_{\mu\nu}^a F^{a,\mu\nu}+ \frac1{2\xi} (\partial A)^2 +
\partial_\mu \bar c^a(D^\mu c)^a\,.
\end{align}
The light-ray operator~\cite{Balitsky:1987bk}  we are interested in  is defined as follows
\begin{equation}
\label{light-ray-def}
    \mathcal{O}(x;z_1, z_2) = \bar{q}(x+z_1n)\left[x+z_1n, x+z_2n\right]\sigma_{\perp +}q(x+z_2 n),
\end{equation}
where $q(x)$ is a quark field, $n$ is an auxiliary light-like ($n^2 = 0$) vector
and
\begin{equation}
    [x+z_1n, x+z_2n] = \operatorname{Pexp}\left\{ ig z_{12}\int_0^1d\alpha\, t^a A^a _+\left(x+z_{21}^\alpha n\right)\right\}
\end{equation}
stands for the Wilson line in the fundamental representation.
Here and below 
\begin{align}
    z_{12}^{\alpha} = z_1\bar{\alpha} + z_2\alpha, && \bar{\alpha} = 1 - \alpha, && z_{12} = z_1 - z_2.
\end{align}
Choosing the second light-like vector $\bar n$ ($(n\bar n)=1$) one expands an arbitrary  $d$-dimensional vector as follows
\begin{align}
a=n (\bar n a) +\bar n (n a) +a_\perp \equiv n a_- + \bar n a_+ +a_\perp,
\end{align}
so that  $\sigma_{\perp +}$  stands for the projection of the matrix
\begin{equation}
\label{sigma-dirac}
    \sigma_{\mu\nu} \equiv \dfrac{1}{2}\left[\gamma_\mu, \gamma_\nu\right]
\end{equation}
 onto the transverse subspace.
In addition, throughout  the paper we omit all the isotopic indices and we use the short-hand notation, $\mathcal{O}(z_1, z_2)$, for the
operator $\mathcal{O}(x=0,z_1, z_2)$.

We also note here that since  $\gamma_\pm$ anticommute with $\gamma_\perp$  the transformation properties of the operator under the
collinear subgroup of the conformal group ($SL(2,\mathbb{R})$ subgroup) are exactly the same as those for the vector operator. Namely,
\begin{align}
\delta_{\pm,0}^\omega\mathcal{O}(z_1, z_2)&= \omega  S_{\pm,0}^{(0)}\mathcal{O}(z_1, z_2),
\end{align}
where $\delta^{\omega}_{\pm,0}$ stand for shifts, dilatations  and special conformal transformations of  a light\nobreakdash-like line and
the corresponding canonical generators  take the form
\begin{align}\label{treelevelgenerators}
S^{(0)}_- =-\partial_{z_1}-\partial_{z_2},
&&
S^{(0)}_0 = z_1\partial_{z_1} +  z_2\partial_{z_2} + 2\,,
&&
S^{(0)}_+ = z_1^2\partial_{z_1}+z_2^2\partial_{z_2}+2z_1+2z_2.
\end{align}

The renormalized operator\,\footnote{Renormalization in
  the modified minimal subtraction scheme ($\overline{\text{MS}}$) will be always tacitly assumed.} is denoted
by $[O](z_1,z_2)$,
\begin{align}\label{ZOrenormalized}
    \left[O\right](z_1, z_2) = Z\mathcal{O}(z_1, z_2),  && Z=\II+\sum_{k>0}\epsilon^{-k} Z_k(a)\,,
\end{align}
where the renormalization factors $Z_k(a)$ are integral operators. The light-ray operator $[O]$ satisfies the RGE
\begin{equation}
\label{ren-rge}
    \left(\mu\dfrac{\partial}{\partial \mu} + \beta(a)\dfrac{\partial}{\partial a} + \mathbb{H}(a)\right)
        \left[\mathcal{O}\right](z_1, z_2) = 0,
\end{equation}
where $\mu$ is the renormalization scale, $a=\alpha_s/4\pi$ is the strong coupling and $\beta(a)$ is the $d$\nobreakdash-dimensional beta
function currently  known with five-loop accuracy~\cite{Baikov:2016tgj,Herzog:2017ohr,Chetyrkin:2017bjc,Luthe:2017ttg}
\begin{align}
\label{beta-func-series}
    \beta(a) = -2a\left(\epsilon + \bar\beta(a)\right)\,, &&\bar\beta(a)=\beta_0 a + \beta_1 a^2 + O(a^3),
\end{align}
with coefficients $\beta_0$, $\beta_1$, etc. in an $SU(N_c)$ gauge theory ($C_F=4/3$, $C_A=N_c=3$ in QCD),
\begin{align}
\label{beta-coefficients}
    \beta_0 =  \dfrac{11}{3}C_A - \frac{2}{3}n_f, && \beta_1 = \dfrac{2}{3}\left(17C_A^2 - 5C_An_f - 3C_Fn_f\right)\,.
\end{align}
The operator $\mathbb{H}(a)$, entering Eq.~\eqref{ren-rge}, is called  the evolution kernel and can be obtained as follows
\begin{equation}
\label{evol-kernel-def}
    \mathbb{H}(a) = -\mu\dfrac{d{Z(a)}}{d\mu}{Z^{-1}(a)} +2\gamma_q(a) = a \mathbb H^{(1)} + a^2 \mathbb H^{(2)} +a^3 \mathbb H^{(3)}+\ldots
    .
\end{equation}
Here $\gamma_q(a)$ is the quark-anomalous dimension and $\mathbb H^{(\ell)}$ are the integral operators of the following type
\begin{equation}
\label{evol-kernl-pert}
    \mathbb{H}^{(\ell)}f(z_1, z_2) = \int_0^1d\alpha \int_0^1d\beta\, h^{(\ell)}(\alpha, \beta)f(z_{12}^\alpha, z_{21}^\beta).
\end{equation}
The one-loop  kernel was obtained in Ref.~\cite{Belitsky:1997rh}. The main purpose of this work is to calculate the two- and three-loop
kernels.

\subsection{Method\label{sect:approach}}
The method of this work fully reflects the approach developed in~\cite{Braun:2016qlg,Braun:2017cih}. The main idea is to consider the
theory in $d=4-2\epsilon$  dimensions at the critical value of the strong coupling $a_*$, such that $\beta(a_*)=0$. Evolution kernels in
the $\overline{\text{MS}}$ scheme do not depend on the space-time dimension and therefore they are essentially the same in the four- and
$d$-dimensional theories. At the critical point  theories enjoy scale and, as a rule, conformal
invariance~\cite{Polyakov:1970xd,Polchinski:1987dy}. This implies that the evolution kernels at the critical point commute with the
corresponding symmetry generators. In the case under consideration these are  generators of the collinear subgroup of the conformal group.
We recall that the tree level generators~\eqref{treelevelgenerators} commute with one-
loop kernel
\begin{equation}
\label{canonical-invariance}
    \left[S^{(0)}_{\pm,0}, \mathbb{H}^{(1)}\right] = 0.
\end{equation}
Beyond one loop the generators receive quantum corrections. Their form is restricted by the requirement for the generators to satisfy the
commutation relations of $\mathrm{sl}(2)$ algebra and give the proper scaling dimensions for local operators
\begin{align}
\label{deformed-generators}
 S_{-} &= S^{(0)}_{-},
 \notag \\
 S_{0} & =S^{(0)}_0 +\Delta S_0= S^{(0)}_0  +\bar \beta(a) + \dfrac{1}{2}\mathbb{H}(a)\,,
\notag \\
 S_{+}  &=S^{(0)}_+  +\Delta S_+ = S^{(0)}_+ + 
(z_1 + z_2)\left(\bar\beta(a)
+ \dfrac{1}{2}\mathbb{H}(a)\right) + z_{12}\Delta(a)\,.
\end{align}
Thus, the corrections to the generators are  expressed in terms of the evolution kernel $\mathbb H(a)$ and an additional operator
$\Delta(a)$ called the conformal anomaly\,\footnote{
We emphasize that there is nothing anomalous in the appearance of this term in the expression for $\mathrm S_+$.
The name ``conformal anomaly'' for the operator $\Delta$ is due to the fact that in scalar field models such a contribution does not arise in low orders.
}.
The conformal  anomaly $\Delta(a)=a\Delta^{(1)} +a^2 \Delta^{(2)}+\ldots$, in lower orders of the perturbation theory can be effectively
extracted from the analysis of the scale and conformal Ward identities for correlators of the light-ray
operators~\cite{Mueller:1993hg,Belitsky:1998gc,Braun:2016qlg}.

Assuming that the conformal anomaly $\Delta(a)$ is known, the invariance of the evolution kernel $\mathbb H(a)$, $[S_+(a), \mathbb
H(a)]=0$, leads to  a chain  of equations\,\footnote{ The kernel $\mathbb  H(a)$ also commutes with the canonical generators $S_-^{(0)} $
and $S_0^{(0)}$. }
\begin{subequations}
\label{orig-equation-set}
\begin{align}
    \left[S_+^{(0)}, \mathbb{H}^{(1)}\right] & = 0\,, \label{SH1}
    \\
    \left[S_+^{(0)}, \mathbb{H}^{(2)}\right] & = \left[\mathbb{H}^{(1)}, \Delta S_+^{(1)}\right]\,,\label{SH2}
     \\
       \label{SH3}
    \left[S_+^{(0)}, \mathbb{H}^{(3)}\right] &=
                                    \left[\mathbb{H}^{(1)}, \Delta S_+^{(2)}\right] + \left[\mathbb{H}^{(2)}, \Delta S_+^{(1)}\right]\,,
\end{align}
\end{subequations}
and so on.
Representing the kernels $\mathbb H^{(\ell)}$ as the sum of canonically invariant and non-invariant parts,
\begin{align}
\mathbb H^{(\ell)}=\mathbb H_\text{inv}^{(\ell)}+\mathbb H^{(\ell)}_\text{non-inv}, && [S^{(0)}_\alpha,\mathbb H_\text{inv}^{(\ell)}]=0,
\end{align}
one sees that Eqs.~\eqref{orig-equation-set} define relations for the non-invariant part of the kernel.
Note that the right hand side of each equation for $\mathbb{H}^{(\ell)}$ involves the kernels of, at most, one order less.  Thus, the knowledge of the
anomaly at order $ \ell-1$ allows us to reconstruct the non-invariant part of the kernel, $\mathbb H^{(\ell)}_\text{non-inv}$, at $\ell$
loops. The invariant part of the evolution kernel, $\mathbb H_\text{inv}^{(\ell)}$, is completely determined by its
eigenvalues, $\gamma_\text{inv}^{(\ell)}(N)= \gamma^{(\ell)}(N) -\gamma_\text{non-inv}^{(\ell)}(N)$, and can be reconstructed in a
relatively simple way, see discussion in Sect.~\ref{Ss-inv}.

\section{Kernel and conformal anomaly}\label{sect:kernels_and_anomaly}
In this section we present explicit expressions for the  evolution kernel and the conformal anomaly at the NLO. We obtained the two-loop
evolution kernel in two ways: by the direct diagram calculation and using the approach described above.
The latter technique is discussed in the next section while the answers for the two-loop diagrams are given in App.~\ref{app:kernel}.

In computing the conformal anomaly we closely follow the approach of Ref.~\cite{Braun:2016qlg}. The operator $\Delta_+$ can be extracted from
the conformal Ward identity for the light-ray operators.
The replacement $\gamma_+\to\sigma_{\perp +}$
in the operator does not affect the analysis given in~\cite[sect.~3]{Braun:2016qlg}. The expression for the operator $\Delta$ in the
first two orders reads~\cite[Eq.(3.47)]{Braun:2016qlg}
\begin{align}
z_{12} \Delta^{(1)} &= z_{12} \Delta^{(1)}_+,
\notag\\
z_{12} \Delta^{(2)} &= z_{12} \Delta^{(2)}_+ + \frac14 \left[\mathbb H^{(2)},z_1+z_2\right].
\end{align}
The operator $\Delta_+$ in the case under consideration can be determined as follows~\cite{Braun:2016qlg}\, \footnote{We present here a
reformulation of  the result of~\cite{Braun:2016qlg} which is more convenient
for practical use.}.
Let us consider the renormalization of the operator $\mathcal O^T(z_1,z_2)$ in QCD perturbed by a local operator,
\begin{align}\label{modifiedQCDaction}
S_{QCD}\mapsto S_\omega=S_{QCD}+ \delta^{\omega}S=S_{QCD} -  2\omega \int d^d y (\bar n y)\left(\frac14 F^2 +\frac1{2\xi}(\partial A)^2\right),
\end{align}
in the leading order in the parameter $\omega$. The renormalized operator takes the form~\eqref{ZOrenormalized} with a modified
renormalization factor,
\begin{align}
Z\mapsto Z_\omega = Z + \omega (n\bar n) \widetilde Z, &&
\widetilde Z =\frac1\epsilon \widetilde Z_1(a)+\frac1{\epsilon^2}\widetilde Z_2+\ldots.
\end{align}
The residues $\widetilde Z_k$ are integral operators and the conformal anomaly is determined by  $\widetilde Z_1$:
\begin{align}
\widetilde Z_1(a)=z_{12}\Delta_+(a)+\frac12 \left[\mathbb H(a)-2\gamma_q(a)\right](z_1+z_2)\,.
\end{align}
We also note  that in the case under consideration there is no mixing with  BRST and EOM operators, see Ref.~\cite{Braun:2018mxm} for
a general analysis.

\subsection{One-loop kernels}\label{sect:oneloop}
\begin{figure}[t]
\centerline{\includegraphics[width=0.65\linewidth]{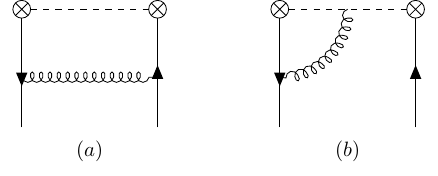}}
\caption{\sf
One-loop Feynman diagrams for the kernel and  the  conformal
anomaly.
}
\label{fig:oneloop}
\end{figure}

The one-loop  diagrams for the kernel are shown in Fig.~\ref{fig:oneloop}. One-loop diagrams for the anomaly have the same topology and
can be obtained from diagrams shown in Fig.~\ref{fig:oneloop} by inserting additional elements generated by $\delta^\omega S$,
cf.~Eq.~\eqref{modifiedQCDaction}. We also note that the exchange diagram (a) in Fig.~\ref{fig:oneloop} does not contribute in both cases due to the gamma matrix identity
\begin{align}
\label{gamma-eps-relation}
    \gamma_\mu \sigma_{\perp +}\gamma^\mu = -2\epsilon \sigma_{\perp +}.
\end{align}

After a short calculation one gets
\begin{subequations}
\begin{align}
\label{one-loop-result}
    \mathbb{H}^{(1)}f(z_1, z_2) &=
    4C_F \left(\int_{0}^1 \frac{d\alpha}\alpha
    \Big(2f(z_1, z_2) - \bar\alpha\big ( f(z_{12}^{\alpha},z_2) + f(z_1, z_{21}^{\alpha})\big )\Big)
    - \frac{3}{2}f(z_1, z_2)\right)
\intertext{and}
\label{one-loop-anomaly}
    \Delta_+^{(1)}f(z_1, z_2) &= -2C_F
    \int\limits_0^1 d\alpha\left(\dfrac{\bar{\alpha}}{\alpha} + \ln\alpha\right)\Big(f(z_{12}^\alpha,z_2) - f(z_1, z_{21}^\alpha)\Big)\,.
\end{align}
\end{subequations}
Let us note that  the one-loop conformal  anomaly~\eqref{one-loop-anomaly}  is exactly the same as  in the  vector
case~\cite{Belitsky:1998gc,Braun:2016qlg}. Calculating the eigenvalues of the kernel $\mathbb H^{(1)}$ by acting on the functions
$\psi_N(z_1,z_2)=z_{12}^{N-1}$ we reproduce the well known
forward anomalous dimensions for the transversity
operators~\cite{Artru:1989zv,Koike:1994st},
\begin{align}
\gamma^{(1)}(N)=4C_F\left[2S_1(N)-\frac32\right].
\end{align}
Here and below $S_{\vec{a}}(N) = S_{a_1,\ldots,a_k}(N)$ stand for the harmonic sums~\cite{Vermaseren:1998uu}.
Our final remark is that one can easily check that the operator $\mathbb H^{(1)}$ commutes, as was expected, with the canonical generators $S_\alpha^{(0)}$.

\subsection{ Two-loop evolution kernel \label{sect:two-loop}}
\begin{figure}[t]
\centerline{\includegraphics[width=0.97\linewidth]{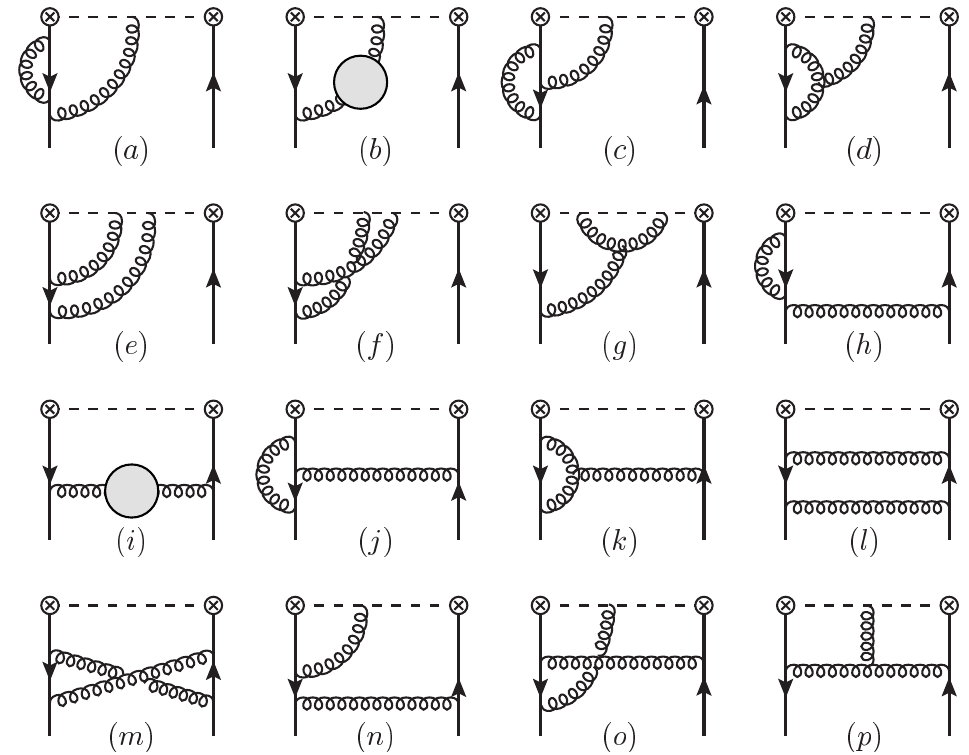}}
\caption{Feynman diagrams of different topology contributing to the two-loop evolution kernel
and the two-loop  conformal anomaly. The grey blob stands for the gluon self-energy insertion. }
\label{fig:two-loop}

\end{figure}

Diagrams contributing to the two-loop evolution kernel are shown in the Fig.~\ref{fig:two-loop}. Answers for the individual diagrams  are
given in the App.~\ref{app:kernel}. Note that the answers for the diagrams without gluon exchange between the quark lines, namely the
diagrams   (a)\,--\,(g) in Fig.~\ref{fig:two-loop}, are exactly the same as in the vector case and we have taken the corresponding results
from~\cite{Braun:2016qlg}.
Contrary, the diagrams (h)\,--\,(p) require separate calculations. Among them, the diagrams (h), (k), (l), (n) do not contribute to the kernel
because of the relation~\eqref{gamma-eps-relation}.

The evolution kernel for the twist-two operators can be written in the following form:
\begin{align}
\label{kernel-structure}
    \mathbb{H}(a) = \Gamma_{\text{cusp}}(a)\widehat{\mathcal {H}}(a) + A(a)  + \mathcal{H}(a).
\end{align}
The first term is completely determined by large $N$ asymptotic of the anomalous dimensions. The kernel $\widehat{ \mathcal H}$ has the
form
\begin{align}
\label{Hhat}
\widehat{\mathcal{H}}f(z_1, z_2) =
\int_0^1\dfrac{d\alpha}{\alpha}\left(2f(z_1, z_2) - \bar{\alpha}\big(f(z_{12}^\alpha, z_2) + f(z_1, z_{21}^\alpha)\big)\right).
\end{align}
It is a canonically invariant operator, $[S^{(0)}_\alpha, \widehat{ \mathcal H}]=0 $, with  eigenvalues, $\widehat{ \mathcal H}
z_{12}^{N-1}=E(N)z_{12}^{N-1}$, equal to $2S_1(N)$. The cusp anomalous dimension, $\Gamma_\text{cusp}(a)$, \cite{Polyakov:1980ca,
Korchemsky:1987wg} is  currently known at four loops~\cite{Henn:2019swt, vonManteuffel:2020vjv}
\begin{align}
\Gamma_\text{cusp}(a) &= a\, 4 C_F +a^2 C_F\left[C_A \left(\frac{268}{9}-8\zeta_2\right) -\frac{40}9 n_f \right]
\notag\\
&\quad
+a^3 C_F\bigg[C_A^2\left(\frac{176}{5}\zeta_2^2+\frac{88}3\zeta_3-\frac{1072}9\zeta_2+\frac{490}{3}\right)
 + C_A n_f\left(-\frac{64}{3}\zeta_3+\frac{160}{9}\zeta_2-\frac{1331}{27}\right)
\notag\\
&\quad
+\frac{n_f}{N_c}\left(-16\zeta_3+\frac{55}{3}\right) -\frac{16}{27}n_f^2 \bigg] + O(a^4)\,.
\end{align}
Next,  $A(a)$ is a constant and  $\mathcal{H}(a)$ is the integral operators of the following form \allowdisplaybreaks{
\begin{align}\label{h-two-loop}
    \mathcal{H}f(z_1, z_2)  =
     &\int_0^1 d\alpha \, \varphi(\alpha)\Big(f(z_{12}^\alpha, z_2) + f(z_1, z_{21}^\alpha)\Big) +
\notag\\
&\quad
    \int_0^1d\alpha\int_0^{\bar{\alpha}} d\beta
    \left(\chi(\alpha, \beta) + \overline \chi(\alpha,\beta)\mathbb P_{12}\right)
    \left(f(z_{12}^\alpha, z_{21}^\beta) + f(z_{12}^\beta, z_{21}^\alpha)\right),
\end{align}
} where the permutation operator $\mathbb P_{12}$ interchanges the variables $z_1, z_2$, i.e.
\begin{align}
  \mathbb P_{12} f(z_1,z_2) = f(z_2,z_1), &&
 \Big( \mathbb P_{12} f(z_{12}^\alpha,z_{21}^\beta) = f(z_{21}^\alpha,z_{12}^\beta) \Big).
\end{align}
The representation~\eqref{kernel-structure} is unique if one supposes that the eigenvalues of the kernel, $\mathcal H(N)$, vanish at $N
\rightarrow\infty$.
Using the results for the diagrams in App.~\ref{app:kernel} we obtain for the constant $A(a)=a A^{(1)} + a^2 A^{(2)}+\ldots $
\begin{align}
A^{(1)} &=-6 C_F\,,
\notag\\
A^{(2)} &= -\frac83 C_F^2\left(\dfrac{43}{8} + 13\zeta_2 \right)
    + 8C_Fn_f\left(\dfrac{1}{12} + \dfrac{2}{3}\zeta_2\right) + \frac{8C_F}{N_c}\left(- \dfrac{17}{24} - \dfrac{11}{3}\zeta_2 + 3\zeta_3 \right),
\end{align}
while for the integral kernels $\varphi, \chi$, and $\overline{\chi}$ we get
\begin{align}  \label{twoloopfunctions}
\varphi^{(2)}(\alpha) &= -4C_F\beta_0 \,\frac{\bar\alpha}{\alpha}\ln\bar\alpha
+ 8 C_F^2 \frac{\bar\alpha}\alpha\ln\bar\alpha\left(\frac32-\ln\bar\alpha + \frac{1+\bar\alpha}{\bar\alpha}\ln\alpha\right),
\notag\\
\chi^{(2)}(\alpha,\beta) &=8 C_F^2\left(\frac1{\bar\alpha}\ln\alpha-\frac1\alpha\ln\bar\alpha  \right)
+\frac{4C_F}{N_c}\left(\frac{\bar\tau}{\tau}\ln\bar\tau +\frac12\right),
\notag\\
\overline \chi^{(2)}(\alpha,\beta) &=\frac{4C_F}{N_c}\left(-\bar\tau \ln\bar\tau +\frac12\right),
\end{align}
where $\tau=\alpha\beta/\bar\alpha\bar\beta$. These expressions are consistent with the result for the two-loop kernel in
momentum fraction representation in refs.~\cite{Belitsky:1998gc, Belitsky:1999gu,Mikhailov:2008my}.

Calculating the forward anomalous dimensions
\begin{align}
    \mathbb{H}(a)z_{12}^{N - 1} = \gamma(N)z_{12}^{N - 1},  && \gamma(N)=a\gamma^{(1)}(N)+a^2\gamma^{(2)}(N)+\ldots
\end{align}
we get  the following expression for $\gamma^{(2)}$  (here and below $S_{\vec{a}}\equiv S_{\vec{a}}(N)$)
\begin{align}\label{gamma-two-loop}
    \gamma^{(2)}(N) &= -8C_F\beta_0\left( S_2 - \frac{5}{3}S_1 +\frac{1}{8}\right)
+8C_F^2\left(-2 S_{2} \left(2S_1-\frac32\right)+\frac{8}3 S_1-\frac{7}{8}\right)
\notag\\
&\quad
+\frac{8 C_F}{N_c}\left(2S_3 - 2S_{-3} + 4S_{1,-2}+ \frac{4}{3}S_1 - \frac{1}{4}
        +\frac{1-(-1)^{N}}{2N(N+1)}\right),
\end{align}
which is in perfect agreement  with the results of Refs.~\cite{ Kumano:1997qp, Hayashigaki:1997dn,Vogelsang:1997ak, Velizhanin:2012nm}.
We have also checked that the kernel $\mathbb H^{(2)}$ satisfies the consistency relation \eqref{SH2}. This implies that although the two-loop
kernel was obtained by direct calculation, it is uniquely determined by the conformal anomaly
$\Delta_+^{(1)}$,~Eq.~\eqref{one-loop-anomaly} and the two-loop anomalous dimensions, Eq.~\eqref{gamma-two-loop}. At present the direct
calculation of the evolution kernel at three loops does not seem to be  feasible, but it can be reconstructed using the two-loop
conformal anomaly and three-loop forward anomalous dimensions.

\subsection{Two-loop anomaly\label{sect:twoloopanomaly}}
The diagrams contributing to the conformal anomaly $\Delta_+$ at two loops can be obtained from the diagrams  shown in
Fig.\,\ref{fig:two-loop} by inserting additional diagrammatic elements
generated by $\delta^{\omega}S$ in Eq.~\eqref{modifiedQCDaction}.
Two such elements are possible: the two-gluon vertex inserted into one of the gluon lines,
or a modified three-gluon vertex replacing the basic three-gluon vertex.
The complete results for the contribution of each Feynman diagram in  Fig.~\ref{fig:two-loop} to the conformal
anomaly  can be found in App.~\ref{app:anomaly}. The technical details and some examples can be found in
Refs.~\cite{Braun:2016qlg,Strohmaier:2018tjo}. We note here that the diagrams without gluon exchange between quark lines, the diagrams
(a)\,--\,(g)  in Fig.~\ref{fig:two-loop}, give rise to the same contribution to $\Delta_+$ as in the vector case.

The kernel $\Delta_+^{(2)}$ can be written in the following form
\begin{align}
\label{Delta+def}
  [\Delta^{(2)}_+ f 
  ](z_1,z_2) &=
  \int_0^1\!du\int_0^1\!dt \, \varkappa(t)\,\Big[f(z_{12}^{ut},z_2) - f(z_1,  z_{21}^{ut})\Big]
\notag\\
&\quad
+\int_0^1\!d\alpha\!\int_0^{\bar\alpha} d\beta \Big[\omega(\alpha,\beta) +\overline\omega(\alpha,\beta) \mathbb{P}_{12}\Big]
\Big[f(z_{12}^{\alpha},z_{21}^\beta) -  f(z_{12}^{\beta},z_{21}^\alpha)\Big].
\end{align}
The function $\varkappa(t)$ is exactly the same as in the vector case, see Refs.~\cite{Braun:2016qlg,Strohmaier:2018tjo}
\begin{align}
\varkappa(t) & =C_F^2\, \varkappa_{{P}}(t)+ \frac{C_F}{N_c} \,\varkappa_{FA}(t)+C_F\beta_0 \varkappa_{bF}(t),
\end{align}
where
\begin{align}\label{varkappa}
\varkappa_{bF}(t) &=
 - 2 \frac{\bar t}{t}
 \Big(  \ln \bar t   + \frac53  \Big),
\notag\\
\varkappa_{FA}(t) & =\frac{2\bar t}{t}\biggl\{ (2+t)\Big[\Li_2(\bar t)-\Li_2(t)\Big]
-(2- t)\Big(\frac{t}{\bar t}\ln t+\ln\bar t\Big)
 - \frac{\pi^2}6 t  -\frac43-\frac{t}{2}\left(1-\frac t{\bar t}\right)\biggr\},
\notag\\
\varkappa_{{P}}(t) & =4\bar t \Big[\Li_2(\bar t)-\Li_2(1)\Big]
+4\left(\frac{t^2}{\bar t}-\frac{2\bar t}t\right)\Big[\Li_2(t)-\Li_2(1)\Big]
-2t\ln t\ln\bar t
-\frac{\bar t}t (2-t) \ln^2\bar t
\notag\\
&\quad
+\frac{t^2}{\bar t}\ln^2 t-2\left(1+\frac1t\right)\ln\bar t
-2\left(1+\frac1{\bar t}\right)\ln t - \frac{16}3\frac{\bar t}t -1-5t\,.
\end{align}
For the functions $\omega$, $\overline\omega$ we obtain
\begin{align}
\overline\omega(\alpha,\beta) &=\frac{C_F}{N_c} \overline\omega_{NP}(\alpha,\beta),
\end{align}
with
\begin{align}
\overline\omega_{NP}(\alpha,\beta) &=
-2\left\{
\frac\alpha{\bar\alpha}\left[\Li_2\left(\frac\alpha{\bar\beta}\right)-\Li_2(\alpha)\right]
  -\alpha\bar\tau\ln\bar\tau-\frac1{\bar\alpha}\ln\bar\alpha\ln\bar\beta-\frac{\beta}{\bar\beta}\ln\bar\alpha -\frac12\beta
\right\}
\end{align}
and
\begin{align}
\omega(\alpha,\beta) &= C_F^2\, \omega_{{P}}(\alpha,\beta)+\frac{C_F}{N_c} \omega_{NP}(\alpha,\beta),
\end{align}
where
\begin{align}
\omega_{{P}}(\alpha,\beta) &=
\frac{4}{\alpha} \Big[\Li_2(\bar \alpha)-\zeta_2+\frac14\bar\alpha\ln^2\bar\alpha+\frac12(\beta-2)\ln\bar\alpha\Big]
\notag\\
&\quad
+\frac{4}{\bar\alpha} \Big[\Li_2( \alpha)-\zeta_2+\frac14\alpha\ln^2\alpha+\frac12(\bar\beta-2)\ln\alpha\Big]\,,
\notag\\[2mm]
\omega_{NP}(\alpha,\beta)&=2
\biggl\{
\frac{\bar\alpha}\alpha\left[\Li_2\left(\frac \beta{\bar\alpha}\right)-\Li_2(\beta)-\Li_2(\alpha)+\Li_2(\bar\alpha)-\zeta_2\right]
-
{\ln\alpha} -\frac1\alpha\ln\bar\alpha
\notag\\
&\quad
+\alpha \left(\frac{\bar\tau}{\tau}\ln\bar\tau+\frac12\right)
\biggr\}\,.
\end{align}
We conclude this section by emphasising that it contains explicit two-loop  expressions of the evolution
kernel~\eqref{kernel-structure} and the conformal anomaly (conformal generators) for the transversity operators~\eqref{Delta+def}.

\section{Three-loop kernel\label{s-three-loop}}

\subsection{Symmetries and kernels}
In this section we explain how to reconstruct  the evolution kernel  from the following data: the forward anomalous dimensions $\gamma(N)$ and
the conformal anomaly $\Delta$. The anomalous dimensions are the eigenvalues of the evolution kernel,
\begin{align}\label{eigenvalues}
\mathbb H(a) z_{12}^{N-1} =\gamma(N)  z_{12}^{N-1}\,.
\end{align}
The kernel $\mathbb H(a)$ is invariant under transformations from the collinear $SL(2,\mathbb{R})$  subgroup of the conformal group
\begin{align}\label{symmetry}
[S_{\pm,0}(a),\mathbb H(a)]=0.
\end{align}
The generators $S_{\pm,0}(a)$ have the form~\eqref{deformed-generators} which includes, besides the evolution kernel itself, the conformal
anomaly $\Delta$.

Although Eqs.~\eqref{eigenvalues} and \eqref{symmetry}, in principle\textcolor{orange}{,} completely determine the  kernel $\mathbb H(a)$,
in practice the problem of finding the kernel is not quite straightforward  since the generators have a non-canonical form.
 To overcome technical problems we  follow the approach developed in
Ref.~\cite{Ji:2023eni} and construct a transformation which maps the deformed symmetry generators to the  canonical ones,
$S_{\pm,0}(a)\mapsto S_{\pm,0}^{(0)}$,
\begin{align}
S_{\pm,0}^{(0)} =\mathrm V S_{\pm,0}(a)\mathrm  V^{-1}\,,  && \mathbf{H}_{\text{inv}}(a) =\mathrm V\, \mathbb H(a)\, \mathrm V^{-1}\,.
\end{align}
The new kernel $\mathbf{H}_{\text{inv}}(a)$ commutes with the canonical  generators, $[S_{\pm,0}^{(0)},\mathbf{H}_{\text{inv}}(a)]=0$, and
has the form
\begin{align}
\label{canonicalkernel}
    \mathbf{H}_{\text{inv}}(a) = \Gamma_{\text{cusp}}(a)\widehat{\mathcal{H}} + \mathcal{A}(a)  + \mathcal{H}(a),
\end{align}
where the kernel $\widehat{\mathcal{H}}$ is defined in Eq.~\eqref{Hhat}, $\mathcal A(a)$ is a constant and
\begin{align}\label{hhbar}
\mathcal{H}(a)f(z_1,z_2)&=\int_0^1d\alpha\int_0^{\bar\alpha}d\beta\left( h(\tau)+\overline h(\tau) \mathbb P_{12}\right)
    f(z_{12}^\alpha,z_{21}^\beta).
\end{align}
The functions $h$ and $\overline h$ are functions of one variable $\tau=\alpha\beta/\bar\alpha\bar\beta$, the so-called conformal ratio.
This property is a consequence of the invariance of the kernel~\eqref{hhbar} under canonical conformal
transformations.\,\footnote{
Note, that the kernel $\mathbf{H}$ which enters Eq.~\eqref{kernel-structure} is parameterized by three functions: a function of one
variable $\varphi(\alpha)$ and two functions of two variables,
$\chi(\alpha,\beta)$ and $\overline\chi(\alpha,\beta)$. Of course, the
invariance of the kernel with respect to the transformations generated by
$S_{\alpha}(a)$ implies some relations between these functions, which, however, are somewhat non-transparent. }
Being a function of one variable, the kernel $h$ ($\overline h$) is completely determined by its moments, $m(N) \ ( \overline{ m}(N))$,
 \begin{align}    \label{mbarm}
m(N) =\int_0^1d\alpha\int_0^{\bar\alpha} d\beta h(\tau) (1-\alpha-\beta)^{N-1} =\int_0^1 \frac{d\tau}{(1-\tau)^2}\, h(\tau)\, Q_N\left(
\frac{1+\tau}{1-\tau}\right)\,,
\end{align}
where $Q_N$ is the Legendre function of the second kind.
Namely,
\begin{align}
h(\tau)=\frac1{2\pi i}\int_C dN \,(2N+1)\, m(N)\, P_N \left(\frac{1+\tau}{1-\tau}\right),
\end{align}
where $P_N$ is the Legendre function of the first kind, and the integration contour $C$ goes along a line parallel to the imaginary axis such that all singularities of $m(N)$ lie to the left of the contour.

\subsection{Similarity transformation\label{sect:similarity}}
The construction of the intertwining operator $\mathrm V$ can be naturally divided into two steps. Let us write, $\mathrm V =\mathrm V_2
\mathrm V_1$. The first transform $\mathrm V_1$ brings the symmetry generators to the ``covariant'' form, $\mathbf S_\alpha(a)=\mathrm V_1
S_\alpha(a) \mathrm V_1^{-1}$,
\begin{align}
\label{generators1}
    \mathbf{S}_{-}(a) & = S^{(0)}_{-},
    \notag\\
    \mathbf{S}_{0}(a)\, & = S_0^{(0)} +\bar \beta(a) + \dfrac{1}{2}\mathbf{H}(a),
    \notag \\
    \mathbf{S}_{+}(a) & = S_+^{(0)} + (z_1 + z_2)\left(\bar\beta(a) + \dfrac{1}{2}\mathbf{H}(a)\right),
\end{align}
where  $\mathbf H(a)=\mathrm V_1 \mathbb H(a) \mathrm V_1^{-1} $. Note that the new generators have the form~\eqref{deformed-generators}
with the conformal anomaly $\Delta(a)\mapsto 0$. An attractive feature of this representation is that when the generators act on an
eigenfunction of the kernel $\mathbf{H}$ one can replace the kernel by the corresponding eigenvalue, namely $\mathbf{H} \mapsto \gamma(N)$.

Looking for the operator $\mathrm V_1$ in the form
\begin{align}
    \mathrm V_1(a) = \exp\big\{\mathrm X(a)\big\},&& \text{where} && \mathrm{X}(a) = a\mathrm{X}^{(1)} + a^2\mathrm{X}^{(2)} + O(a^3),
\end{align}
one gets the following equations for  $\mathrm{X}^{(k)}$:
\begin{equation}
\label{x-pm-inv}
    [S_-^{(0)},\mathrm X^{(k)}]=[S_0^{(0)},\mathrm X^{(k)}]=0
\end{equation}
and
\begin{subequations}
\label{x-equations}
\begin{align}
    \left[S_{+}^{(0)}, \mathrm{X}^{(1)}\right] &= z_{12}\Delta^{(1)},
    \label{Xeq-1}
     \\
    \left[S_{+}^{(0)}, \mathrm{X}^{(2)}\right] &= z_{12}\Delta^{(2)}
    + \left[\mathrm{X}^{(1)}, z_1 + z_2\right]\left(\beta_0 + \frac{1}{2}\mathbb{H}^{(1)}\right)
     + \dfrac{1}{2}\left[\mathrm{X}^{(1)}, z_{12}\Delta^{(1)}\right].
\label{Xeq-2}
\end{align}
\end{subequations}
These equations define the operators $\mathrm X^{(k)}$ up to a canonically invariant operator. It reflects the arbitrariness in the
definition of $\mathrm V_1$, which can be multiplied  by an arbitrary operator depending only on the kernel $\mathbf{H}$:
\begin{align}
\mathrm V_1\mapsto \mathrm V_1^\prime= \mathrm U(\mathbf{H}) \mathrm V_1.
\end{align}
Since the relation~\eqref{x-pm-inv} holds, the operators $\mathrm X^{(k)}$ can be represented as  integral operators similar
to~\eqref{h-two-loop}. The Eqs.~\eqref{x-equations} lead to differential equations on the integral kernels which are not difficult
to solve. For example, the operator $\mathrm X^{(1)}$ has the form
\begin{align}\label{X1}
    \mathrm X^{(1)}f(z_1, z_2) = 2C_F\int\limits_0^1 d\alpha\,
    \frac{\operatorname{ln}\alpha}{\alpha}\Big(2f(z_1,z_2) - f(z_{12}^{\alpha}, z_2)  - f(z_1, z_{21}^{\alpha})\Big),
\end{align}
which is exactly the same as in the vector case. The expression for the kernel $\mathrm X^{(2)}$ is quite involved and is given  in
App.~\ref{sect:X}, while  we move to the second transformation,~$\mathrm V_2$. Remarkably enough it can be written in a closed
form~\cite{Ji:2023eni}
 \begin{align}
 \label{v2-closed-form}
\mathrm V_2  =\sum_{k=0}^\infty \frac1{k!} \mathrm L^k \left(\bar\beta(a)+\frac12 \mathbf{H}(a)\right)^k\,,
&&
\mathrm V_2^{-1} =\sum_{k=0}^\infty \frac1{k!} (-\mathrm L)^k \left(\bar\beta(a)+\frac12 \mathbf{H}_{\text{inv}}(a)\right)^k\,,
\end{align}
where $\mathrm L= \ln |z_{12}|$.

The operator $\mathrm V_2$ intertwines the generators~\eqref{generators1} with the canonical ones  and the kernels $\mathbf H$ and
$\mathbf{H}_{\text{inv}}$
\begin{align}
\mathrm V_2\,\mathbf{S}_\alpha(a) =  S_\alpha^{(0)}\, \mathrm V_2,  && \mathrm V_2\,\mathbf{H}(a) =  \mathbf{H}_{\text{inv}}(a)\, \mathrm V_2\,.
\end{align}
Inserting~\eqref{v2-closed-form} in the last of these equations we obtain the following relation between the kernels $\mathbf H$ and
$\mathbf H_\text{inv}$,
\begin{align}\label{HH}
\mathbf{H}(a) =\mathbf{H}_{\text{inv}}(a)+\sum_{n=1}^\infty \frac1{n!} \mathrm T_n(a)\,\left(\bar\beta(a)+\frac12\mathbf{H}
(a)\right)^n\,,
\end{align}
where the operators $\mathrm T_n(a)$ are defined by recursion,
\begin{align}
\mathrm T_n(a)=[\mathrm T_{n-1}(a),\mathrm L]\,, && \mathrm T_0(a)=\mathbf{H}_{\text{inv}}(a)\,.
\end{align}
Taking into account Eqs.~\eqref{canonicalkernel}, \eqref{hhbar} one gets for $\mathrm T_n(a)$,  $n>0$,
\begin{align}
\mathrm T_n(a) f(z_1,z_2) &= -\Gamma_\text{cusp}(a) \int_0^1d\alpha\frac{\bar\alpha}\alpha\ln^n\bar\alpha
\left(f(z_{12}^\alpha,z_2)+ f(z_1,z_{21}^\alpha)\right)
\notag\\
&\quad
+\int_0^1d\alpha\int_0^{\bar\alpha}d\beta \ln^n(1-\alpha-\beta)
\Big(h(\tau)+\overline h(\tau)\mathbb P_{12}\Big) f(z_{12}^\alpha, z_{21}^\beta)\,.
\end{align}
Since the $n$-th term in the sum in~\eqref{HH} is of order $O(a^{n+1})$ one can easily obtain an   approximation for $\mathbf H(a)$ with
arbitrary precision, e.g.
\begin{align}
\mathbf{H}(a) & = \mathbf{H}_{\text{inv}}(a)+ \mathrm T_1(a) \left(1+\frac12 \mathrm T_1(a) \right)
\left(\bar \beta(a)+\frac12 \mathbf{H}_{\text{inv}}(a)\right)
 \notag \\
&\quad
+\frac12 \mathrm T_2(a)\left(\bar \beta(a)+\frac12 \mathbf{H}_{\text{inv}}(a)\right)^2 +O(a^4)\,.
\end{align}
 Expanding  all operators in power series,
  $\mathbf{H}_{\text{inv}}(a)=\sum_k a^k \mathbf{H}_{\text{inv}}^{(k)}$, $\mathrm T_n(a)=\sum_k a^k \mathrm T_n^{(k)}$, one
derives
\begin{subequations}
\label{boldface-solution}
\begin{align}
\label{bfH1}
    \mathbf H^{(1)} &= \mathbf{H}_{\text{inv}}^{(1)},
     \\
     \label{bfH2}
    \mathbf H^{(2)} &= \mathbf{H}_{\text{inv}}^{(2)} + \mathrm T_1^{(1)}\left(\beta_0
     + \dfrac{1}{2}\mathbf{H}_{\text{inv}}^{(1)}\right),
      \\
      \label{bf3}
    \mathbf H^{(3)} &= \mathbf{H}_{\text{inv}}^{(3)}
    + \mathrm T_1^{(1)}\!\left(\beta_1\!+\! \frac{1}{2}\mathbf{H}_{\text{inv}}^{(2)}\right)
    + \frac12 \mathrm T_2^{(1)}\!\left(\beta_0\! +\! \frac{1}{2}\mathbf{H}_{\text{inv}}^{(1)}\right)^2\!\!
    + \left(\mathrm T_1^{(2)}\! + \! \frac{1}{2}\left(\mathrm T_1^{(1)}\right)^2\right)
       \!\left(\beta_0\! +\! \frac{1}{2}\mathbf{H}_{\text{inv}}^{(1)}\right),
\end{align}
\end{subequations}
which agrees with the expressions obtained in Refs.~\cite{Braun:2017cih,Strohmaier:2018tjo}.

Concluding this section we discuss the relation between the eigenvalues of the operators $\mathbf H$ and $\mathbf H_\text{inv}$.
Since both operators commute with the permutation operator $\mathbb P_{12}$,  functions symmetric and anti-symmetric under permutations
$z_1\leftrightarrow z_2$ form invariant subspaces of both operators. It is easy to check that the functions $\psi_N^+(z_1,z_2)=
|z_{12}|^{N-1}$ and $\psi_N^-(z_1,z_2)=\text{sign}(z_{12}) |z_{12}|^{N-1}$ are the eigenfunctions of both operators. Note that we do not
assume that $N$ is integer. Then if
\begin{align}
\mathbf{H}(a)\psi^\pm_N(z_1,z_2)= \gamma_\pm (N)\psi^\pm_N(z_1,z_2), &&\text{and}&&
\mathbf{H}_{\text{inv}}(a)\psi^\pm_N(z_1,z_2)= \lambda_\pm (N)\psi^\pm_N(z_1,z_2),
\end{align}
using the relation~\eqref{HH}, one gets the following relation for the eigenvalues of $\gamma_\pm$ and $\lambda_\pm$
\begin{align}
\gamma_\pm(N) = \lambda_\pm \left(N+\bar\beta(a)+\frac12\gamma_\pm(N)\right).
\end{align}
This relation was introduced in Refs.~\cite{Dokshitzer:2005bf,Basso:2006nk} as a generalization of the Gribov-Lipatov reciprocity
relation~\cite{Gribov:1972ri,Gribov:1972rt}. The functions $\lambda_\pm$ have much simpler form than the anomalous dimensions~$\gamma_\pm$.
The asymptotic expansion of the functions $\lambda_\pm(N)$ for large $N$ is invariant under the reflection $N\to -N-1$, see
e.g.~\cite{Basso:2006nk,Dokshitzer:2006nm,Beccaria:2009vt,Alday:2015eya}. This means that only special combinations of the harmonic
sums~\cite{Remiddi:1999ew} can appear in the perturbative expansion of reciprocity respecting (RR) anomalous
dimensions~\cite{Beccaria:2009vt}. Thus starting from the three loop anomalous dimensions for the transversity
operators~\footnote{
The three loop anomalous dimensions for general $N$ were  reconstructed from the first $15$ moments in ref.~\cite{Velizhanin:2012nm}.
This result was later confirmed by the direct calculation~\cite{Blumlein:2021enk}.
}~\cite{Velizhanin:2012nm,Blumlein:2021enk} we can find the RR anomalous dimensions, $\lambda_\pm(N)$, and, using the technique developed
in~\cite{Ji:2023eni}, reconstruct the kernel~$\mathbf H_\text{inv}$. Then the kernels $\mathbf H^{(k=1,2,3)}$ are given by
Eqs.~\eqref{boldface-solution} and the evolution kernels in $\overline{\text{MS}}-$scheme read,
\begin{subequations}
\label{Hmsbar}
\begin{align}
\label{Hms1}
\mathbb H^{(1)} &= \mathbf H^{(1)}\,,
\\
\label{Hms2}
\mathbb H^{(2)} &= \mathbf H^{(2)} +[\mathbf H^{(1)},\mathrm X^{(1)}]\,,
\\
\label{Hms3}
\mathbb H^{(3)} &= \mathbf H^{(3)} + [\mathbf H^{(2)},\mathrm X^{(1)}] + [\mathbf H^{(1)},\mathrm X^{(2)}]
+\frac12 [[\mathbf H^{(1)},\mathrm X^{(1)}]\,\mathrm X^{(1)}]\,.
\end{align}
\end{subequations}
The kernel $\mathrm{X}^{(1)}$ is presented in~\eqref{X1} and the explicit expression for the  kernel $\mathrm X^{(2)}$ can be found in
App.~\ref{sect:X}.

\subsection{Invariant kernel\label{Ss-inv}}
The kernels $h$, $\overline h$  which determine the operator $\mathcal H(a)$ in Eq.~\eqref{hhbar} can be obtained as follows: First, we
reconstruct the eigenvalues of the kernel $\mathbf H_\text{inv}$,  $\lambda_\pm(N)$, using the result for the three loop anomalous
dimensions $\gamma_\pm(N)$, \cite{Velizhanin:2012nm, Bagaev:2012bw}. The above mentioned functions can be written as
\begin{align}
\gamma_\pm(N) & =2\Gamma_\text{cusp}(a) S_1(N)+ A(a) + {\kappa_\pm}(N),
\notag\\[2mm]
\lambda_\pm(N)& =2\Gamma_\text{cusp}(a) S_1(N)+ \mathcal A(a) + {m_\pm}(N).
\end{align}
The anomalous dimensions $\gamma_+$ and $\gamma_-$ gives the anomalous dimensions of the local operators for even and odd $N$, respectively,
and  $m_\pm(N)=m(N)\mp \overline{m}(N)$, where $m(N),\overline{m}(N)$ are the moments of the kernels $h,\bar h$, Eqs.~\eqref{mbarm}.
In the leading order  $m^{(1)}_\pm(N)=0$ and $\mathcal A^{(1)} 
= -6C_F$. At two loops one finds
\begin{align}
\mathcal A^{(2)}&=-C_F^2\left( \frac{43}{3} + \frac{104}3 \zeta_2 \right) + C_F n_f \left( \frac23+ \frac{16}3\zeta_2 \right)
+ \frac{ C_F}{N_c}\left( - \frac{17}3 - \frac{88}3\zeta_2 + 24\zeta_3\right),
\notag\\[2mm]
m^{(2)}_\pm(N) &= \frac{2 C_F}{N_c}\Biggl(32 S_1 \left(S_{-2}+\frac{\zeta_2}2\right) +16 \left(S_{3}-\zeta_3\right)
-32\left(S_{-2,1}-\frac12S_{-3}+\frac13\zeta_3\right)
\notag\\
&\quad+ \frac{2\left(1-(-1)^N\right)}{N(N+1)}\Biggr),
\end{align}
The expression for the moments $m_\pm$ includes only  special combinations of  harmonic
sums, the so-called parity invariant harmonic sums~\cite{Beccaria:2009vt}, whose asymptotic expansion is invariant under $N\mapsto-N-1$.
Namely, following~\cite{Ji:2023eni}, we define
\begin{align}
\Omega_1(N)&=S_1(N),   &&
\Omega_{-2}(N)=(-1)^N\left(S_{-2}(N)+\frac{\zeta_2}2\right),
\notag\\
\Omega_3(N) &=S_3(N)-\zeta_3, &&
\Omega_{-2,1}(N)=(-1)^N\left(S_{-2,1}(N)-\frac12S_{-3}(N)+\frac13\zeta_3\right),
\end{align}
and rewrite $ m_\pm$ as
\begin{align}
m^{(2)}_\pm(N) &= \frac{2 C_F}{N_c}\left( 16 \Omega_3 \pm 32\left(\Omega_1 \Omega_{-2}+\Omega_{-2,1}\right) +\frac{2(1\mp 1)}{N(N+1)}\right).
\end{align}
The kernels with eigenvalues corresponding to $\Omega_{a,b,\ldots}$ can be effectively constructed, see ~\cite{Ji:2023eni}, e.g.
$\Omega_{-2}\mapsto \bar\tau/2$, $\Omega_3\mapsto \bar\tau/(2\tau)\ln\bar\tau$, see App.~\ref{sect:Omegas}. The product of two sums
$\Omega_{\vec{a}}\times \Omega_{\vec{b}}$ corresponds to the convolution of the corresponding  kernels that can be easily evaluated with
the \textsc{HyperInt} package~\cite{Panzer:2014caa}.  Thus, after some algebra, we obtain for the kernels $h,\overline{h}$
\begin{align}
h^{(2)}(\tau) = \frac{8 C_F}{N_c}\left( \frac{\bar\tau}\tau \ln\bar\tau +\frac12\right),
&&
\overline{h}^{(2)}(\tau) = \frac{8 C_F}{N_c}\left(-{\bar\tau}\ln\bar\tau +\frac12\right),
\end{align}
which is in full agreement with the result of the explicit calculation, Eq.~\eqref{twoloopfunctions}.

Going to the three-loop expression and repeating all the steps described above we obtain
\begin{align}
\mathcal A^{(3)}&=
C_F\,n_f^2\left(\frac{34}9 -\frac{160}{27}\zeta_2 + \frac{32}9\zeta_3\right)
+
C_F^2\,n_f\left(-34 + \frac{4984}{27}\zeta_2-\frac{512}{15}\zeta_2^2 + \frac{16}9 \zeta_3\right)
\notag\\
&\quad
+\frac{C_F n_f}{N_c}\left(-40+\frac{2672}{27}\zeta_2 - \frac 8 5 \zeta_2^2 - \frac{400}{9}\zeta_3 \right)
\notag\\
&\quad
+C_F^3\left(\frac{1694}9 - \frac{22180}{27}\zeta_2
+ \frac{2464}{15}\zeta_2^2 + \frac{1064}9\zeta_3 -  320\zeta_5\right)
\notag\\
&\quad
+\frac{C_F^2}{N_c}\left(\frac{5269}{18} - \frac{28588}{27}\zeta_2 + \frac{2216}{15}\zeta_2^2+\frac{7352}9\zeta_3 - 32\zeta_2\zeta_3
-560\zeta_5\right)
\notag\\
&\quad
+\frac{C_F}{N_c^2}\left(
\frac{1657}{18} - \frac{8992}{27}\zeta_2 + 4\zeta_2^2 + \frac{3104}9\zeta_3 - 80\zeta_5
\right).
%
\end{align}
%
For the three-loop kernels $h^{(3)}$ and $\overline{h}^{(3)} $ we find
\allowdisplaybreaks{
\begin{subequations}
\begin{align}\label{three-loop-int-kernel}
h^{(3)}(\tau)
  &= -C_F\* n_f^2\frac{16}{9}
+ C_F^2\* n_f\left( \frac{352}{9} -\frac{8}{3} \mathrm{H}_0 +\frac{16}{3}\frac{\bar\tau}{\tau}\left( \mathrm{H}_2- \mathrm{H}_{10}\right) \right)
\notag\\
&\quad
+\frac{C_F n_f} {N_c} \left( 8 -\frac{8}{3} \mathrm{H}_1 -\frac{4}{3} \mathrm{H}_0
+ \frac{\bar\tau}{\tau}\left( 8 \mathrm{H}_2-\frac{8}{3} \mathrm{H}_{10}+\frac{16}{3} \mathrm{H}_{11} +\frac{160}{9} \mathrm{H}_1 \right)
\right)
\notag\\
&\quad
+ C_F^3\left( -\frac{1936}{9} +\frac{88}{3} \mathrm{H}_0 + 32\frac{\bar\tau}{\tau}\*
\left( \mathrm{H}_{3} + \mathrm{H}_{12} - \mathrm{H}_{110}- \mathrm{H}_{20} -\frac{1}{3} \mathrm{H}_2 +\frac{1}{3} \mathrm{H}_{10} +\frac{1}{2} \mathrm{H}_1\right)
\right)
\notag\\
&\quad
+\frac{C_F^2}{N_c}\Bigg(
-\dfrac{152}{3} - 96\zeta_3 - \bigg(\frac{8}{3}
- 48\zeta_2\bigg)\mathrm{H}_0 +\frac{76}{3}\mathrm{H}_{1} -32\mathrm{H}_{10} + 4\mathrm{H}_{2}
- 48\mathrm{H}_{20} - 16\mathrm{H}_{11}
\notag \\
&\quad   - 24\mathrm{H}_{21} + \dfrac{\tau}{\bar{\tau}}\bigg(- 24\zeta_2 - 48\zeta_3 + 64\mathrm{H}_{0}\bigg)
+ \dfrac{\tau + 1}{\bar{\tau}}\bigg( -\big(32 - 16\zeta_2\big)\mathrm{H}_0
\notag\\
&\quad
+12\mathrm{H}_2 - 16\mathrm{H}_{20} - 8\mathrm{H}_{21}\bigg)
+ \dfrac{\bar{\tau}}{\tau}\bigg(-\bigg(\frac{2000}{9} + 16\zeta_2\bigg)\mathrm{H}_1 + \frac{32}{3}\mathrm{H}_{10} -\frac{208}{3}\mathrm{H}_2
\notag \\
&\quad
- 64\mathrm{H}_{20} - \dfrac{32}{3}\mathrm{H}_{11} - 32\mathrm{H}_{110} + 64\mathrm{H}_3  + 80\mathrm{H}_{12}
 + 64\mathrm{H}_{21} + 96\mathrm{H}_{111}\bigg)\Bigg)
\notag\\
&\quad
+\frac{C_F}{N_c^2}\*
\Bigg(\frac{544}{9} + 16\zeta_2 - 96\zeta_3 - \bigg(\frac{68}{3} - 36\zeta_2\bigg)\mathrm{H}_{0}
+ \frac{68}{3}\mathrm{H}_1 - 24\mathrm{H}_{10} + 4\mathrm{H}_2 -36\mathrm{H}_{20}
\notag\\
&\quad
 + \dfrac{\tau}{\bar{\tau}}\bigg(-8\zeta_2 - 48\zeta_3 + 48\mathrm{H}_0 \bigg)
+ \dfrac{\tau + 1}{\bar{\tau}}\bigg(\big(-24 + 12\zeta_2\big)\mathrm{H}_0 + 4\mathrm{H}_2 -12\mathrm{H}_{20} \bigg)
\notag\\
&\quad
+\dfrac{\bar{\tau}}{\tau}\bigg(-\bigg(\frac{1072}{9} + 16\zeta_2\bigg)\mathrm{H}_1
+ \frac{44}{3}\mathrm{H}_{10} - 44\mathrm{H}_{2} - 32\mathrm{H}_{20} -\frac{16}{3}\mathrm{H}_{11} - 16\mathrm{H}_{110}
\notag\\
&\quad + 32\mathrm{H}_3 + 32\mathrm{H}_{12} + 48\mathrm{H}_{21} + 32\mathrm{H}_{111}\bigg)\Bigg)\,,
\\[2mm]
\overline{h}^{(3)}(\tau)
  & =-\frac{C_F n_f} {N_c}  \left( \frac{104}{9} +\frac{8}{3} \mathrm{H}_0 +\frac{8}{9} \Big(23-20\tau\Big) \mathrm{H}_1
+\frac{16}{3}\bar{\tau}\Big( \mathrm{H}_{11}+ \mathrm{H}_{10}\Big)
\right)
\notag\\
&\quad
+\frac{C_F^2}{N_c}\*
\Bigg(\dfrac{1480}{9} - 40\zeta_2 - 48\zeta_3  +\bigg(\frac{28}{3} + 24\zeta_2\bigg)\mathrm{H}_0
+ \frac{76}{3}\mathrm{H}_1 + 16\mathrm{H}_{10} - 4\mathrm{H}_2 - 24\mathrm{H}_{20}
\notag\\
&\quad
- 16\mathrm{H}_{11} + 24\mathrm{H}_{21} + \dfrac{\tau}{\bar{\tau}}\bigg(-24\zeta_2 + 48\zeta_3 - 32\mathrm{H}_0\bigg)
+ \dfrac{\tau + 1}{\bar{\tau}}\bigg(\bigg(16 - 8\zeta_2\bigg)\mathrm{H}_0 + 12\mathrm{H}_2
\notag\\
&\quad
 + 8\mathrm{H}_{20} - 8\mathrm{H}_{21}\bigg) + \bar{\tau}\bigg(-24 + 48\zeta_2 + 48\zeta_3 - 16\zeta_2\mathrm{H}_0
+ \bigg(\frac{2144}{9} + 16\zeta_2\bigg)\mathrm{H}_1 + \frac{104}{3}\mathrm{H}_{10}
\notag \\
&\quad
 - 24\mathrm{H}_2 + 16\mathrm{H}_{20}
+ \frac{32}{3}\mathrm{H}_{11}
- 16\mathrm{H}_{110} -32\mathrm{H}_{12} - 32\mathrm{H}_{21} - 96\mathrm{H}_{111}\bigg)\Bigg)
\notag\\
&\quad
+\frac{C_F}{N_c^2}\*
\Bigg( \dfrac{1028}{9} - 24\zeta_2 - 48\zeta_3 + \bigg(\frac{44}{3} + 36\zeta_2\bigg)\mathrm{H}_0
+ \frac{68}{3}\mathrm{H}_1 + 24\mathrm{H}_{10} - 4\mathrm{H}_2 - 36\mathrm{H}_{20}
\notag\\
&\quad
+ \frac{\tau}{\bar{\tau}}\bigg(-8\zeta_2 + 48\zeta_3 - 48\mathrm{H}_{0}\bigg)
 + \dfrac{\tau + 1}{\bar{\tau}}\bigg(\bigg(24 - 12\zeta_2\bigg)\mathrm{H}_0 + 4\mathrm{H}_2 + 12\mathrm{H}_{20}\bigg)
\notag\\
&\quad
+\bar{\tau}\bigg(-24 + 24\zeta_2 + 48\zeta_3 - 32\zeta_2\mathrm{H}_0
  + \bigg(\frac{1072}{9}+ 16\zeta_2\bigg)\mathrm{H}_1  + \frac{88}{3}\mathrm{H}_{10}
\notag  \\
 &\quad
  - 24\mathrm{H}_2 + 32\mathrm{H}_{20} + \frac{16}{3}\mathrm{H}_{11} - 32\mathrm{H}_{110} + 16\mathrm{H}_{12}
+ 16\mathrm{H}_{21} - 32\mathrm{H}_{111}\bigg)\Bigg)
   ,
\end{align}
\end{subequations}
where $\mathrm{H}_{\vec{a}}(\tau) \equiv \mathrm{H}_{a_1 \ldots a_k}$ are harmonic polylogarithms (HPLs)~\cite{Remiddi:1999ew}.
}

\section{Local operators} \label{sect:local}
In this section we present the anomalous dimension matrix for the local operators in the Gegenbauer basis,
\begin{align}\label{localOnk}
\mathcal O_{nk}(0) =(\partial_{z_1}+\partial_{z_2})^k
C^{(3/2)}_n \left(
\frac{ \partial_{z_1}-\partial_{z_2} }{ \partial_{z_1}+\partial_{z_2}}\right)[\mathcal O](z_1,z_2)\Big|_{z_1=z_2=0} \,,
\end{align}
where $k\geq n$ are integers. The RGE for these operators takes the form
\begin{align}
\left( \mu\frac{\partial}{\partial\mu} + \beta(a)\frac{\partial}{\partial a} \right) O_{nk} =-\sum_{n'=0}^n\gamma_{nn'} O_{n'k}\,.
\end{align}
Note that the anomalous dimension matrix does not  depend on $k$. In the Gegenbauer basis the matrix $\gamma_{nn'}$ is diagonal at one loop
\begin{align}
\gamma_{nn'}^{(1)} & = \delta_{nn'} \gamma^{(1)}(n+1) =\delta_{nn'} 4 C_F\left[2S_1(n+1)-\frac32  \right]\,,
\end{align}
i.e. the operators $\mathcal O_{nk}$ evolve autonomously in this order~\cite{Makeenko:1980bh}. It easy to understand that the anomalous
dimension matrix $\gamma$ is nothing else as a matrix of the evolution kernel $\mathbb H$ in a certain basis. See, e.g.,
Ref.~\cite{Moch:2021cdq,VanThurenhout:2023gmo} for a discussion of their basis transformation properties. Indeed, expanding the light-ray
operator over the local operators as follows
\begin{align}
[\mathcal O](z_1,z_2)=\sum_{kn} \Psi_{nk}(z_1,z_2) \mathcal O_{nk}(0) \,,
\end{align}
one defines the functions $\Psi_{nk}(z_1,z_2) $, which are homogeneous polynomials of degree $k$ in $z_1,z_2$, e.g. $\Psi_{nk}(z_1,z_2)\sim
(S_+^{(0)})^{k-n}(z_1-z_2)^n$.
These functions diagonalize the one-loop kernel and beyond one loop one obtains
\begin{align}
\mathbb H \Psi_{nk} = \sum_{n'=0}^n\gamma_{n'n} \Psi_{n'k}\,.
\end{align}
Thus the off-diagonal part of the anomalous dimension matrix $\gamma$ is completely determined by the non-invariant part of the kernel.
Namely, evaluating Eqs.~\eqref{orig-equation-set} in the basis formed by the functions  $\Psi_{nk}$ one can easily reconstruct the
off-diagonal part of the matrix $\gamma$. The method was developed by Dieter M\"uller in \cite{Mueller:1991gd},  while here we follow an
analysis given in Ref.~\cite{Braun:2017cih}.

At two loops the off-diagonal part of the anomalous dimension matrix  can be written in analytical form:
\begin{align}\label{gamma2off}
 \gamma^{(2)}_{mn} &= \delta_{mn}\gamma^{(2)}_n - \frac{\gamma^{(1)}_m-\gamma^{(1)}_n}{a_{mn}}
\biggl\{
 - 2(2n+3) \left(\beta_0 + \frac12 \gamma^{(1)}_n\right)\vartheta_{mn}+  { w}_{mn}^{(1)}
\biggr\},
\end{align}
where $\gamma_n\equiv\gamma_{nn}$,
\begin{align}
a_{mn}& =(m-n)(m+n+3)\,,
\notag\\
 {w}_{mn}^{(1)} &= 4 C_F (2n+3)\, {a}_{mn} \left( \frac{A_{mn} - S_1(m+1)}{(n+1)(n+2)}
  + \frac{2A_{mn}}{a_{mn}} \right) \vartheta_{mn}\,,
\notag\\
A_{mn} &= S_1\left(\frac{m+n+2}2\right) -S_1\left(\frac{m-n-2}2\right) +2S_1(m-n-1)-S_1(m+1).
\end{align}
and
\begin{align*}
  \vartheta_{mn} =
\begin{cases}
1 \text{ if } m - n > 0 \text{ and even} \\ 0 \text { else.}
\end{cases}
\end{align*}
The Eq.~\eqref{gamma2off} is the same as in the vector case~\cite{Belitsky:1998gc}. Of course, one can take the corresponding
diagonal anomalous dimension $\gamma_n$. For the first few elements of the matrix we obtained (for $N_c=3$):
\begin{align}
\gamma^{(2)} &= \begin{pmatrix}
\frac{724}9 & 0 & 0 & 0 & 0 & 0
\\
0 & 124 & 0 & 0 & 0 & 0
\\
\frac{272}9& 0& \frac{38044}{243} &0 &0 &0
\\
0 &  \frac{8360}{243} & 0& \frac{44116}{243} & 0 & 0
\\
\frac{44}{5} & 0 &  \frac{4592}{135}& 0 & \frac{6155756}{30375} & 0
\\
0 & \frac{5852}{405} & 0 & \frac{36512}{1125} &  0 & \frac{744184}{3375}
\end{pmatrix}
%
%
%
%
%
%
-n_f \begin{pmatrix}
\frac{104}{27}& 0 & 0 & 0 & 0 & 0
\\
0 & 8 & 0 & 0 & 0 & 0
\\
\frac{32}9& 0& \frac{904}{81} & 0& 0& 0
\\
0 &  \frac{80}{27} & 0& \frac{1108}{81} & 0 &0
\\
\frac{88}{45} & 0 & \frac{112}{45}& 0& \frac{31924}{2025} & 0
\\
0 & \frac{152}{81} & 0 & \frac{32}{15} & 0& \frac{35524}{2025}
\end{pmatrix}\,.
\end{align}
For the three-loop matrix $\gamma^{(3)}$ there is no analytical expression.
As above we give the numerical expression for the first few off-diagonal elements, ($0\leq m,n\leq 5$) for $N_c=3$,
\begin{align}
\gamma^{(3)}_\text{off}=\gamma^{(3)}_1 + n_f \gamma^{(3)}_{n_f}+ n_f^2 \gamma^{(3)}_{n_f^2}\,.
\end{align}
We find
\begin{align}
\gamma^{(3)}_1 &= \begin{pmatrix}
0& 0 & 0 & 0 & 0 & 0
\\
0& 0 & 0& 0 & 0 & 0
\\
 \frac{44992}{81}& 0& 0 & 0& 0& 0
\\
0 &  \frac{ 1316680} {2187} & 0&  0 & 0 &0
\\
\frac{1977808}{10125} & 0 &   \frac{54669748}{91125}& 0&  0 & 0
\\
0 & \frac{68848018}{273375}& 0 &  \frac{443231668}{759375} & 0& 0
\end{pmatrix}
\end{align}
and
\begin{align}
\gamma^{(3)}_{n_f} = -\begin{pmatrix}
0& 0 & 0 & 0 & 0 & 0
\\
0& 0 & 0& 0 & 0 & 0
\\
 \frac{21008}{243}& 0& 0 & 0& 0& 0
\\
0 &  \frac{200060}{2187} & 0&  0 & 0 &0
\\
\frac{998842}{30375} & 0 &   \frac{898436}{10125}& 0&  0 & 0
\\
0 & \frac{745418}{18225}& 0 &  \frac{4266496}{50625} & 0& 0
\end{pmatrix},
&&
%
\gamma^{(3)}_{n_f^2} = -\begin{pmatrix}
0& 0 & 0 & 0 & 0 & 0
\\
0& 0 & 0& 0 & 0 & 0
\\
 \frac{160}{81}& 0& 0 & 0& 0& 0
\\
0 &  \frac{520}{243} & 0&  0 & 0 &0
\\
\frac{1012}{2025} & 0 &   \frac{4088}{2025}& 0&  0 & 0
\\
0 & \frac{3268}{3645}& 0 &  \frac{416}{225} & 0& 0
\end{pmatrix}.
\end{align}
For completeness we also provide the first few diagonal entries of the anomalous
dimension,
\begin{align}
\gamma_{00}^{(3)} &= \frac{105110}{81} - \frac{1856}{27}\zeta_3 - \left(\frac{10480}{81} + \frac{320}{9}\zeta_3\right)n_f
- \frac{8}{9}n_f^2,
\nonumber\\
\gamma_{11}^{(3)} &= \frac{19162}{9}  - \left(\frac{5608}{27} + \frac{320}{3}\zeta_3\right)n_f - \frac{184}{81}n_f^2,
\nonumber\\
\gamma_{22}^{(3)} &= \frac{17770162}{6561} + \frac{1280}{81}\zeta_3 - \left(\frac{552308}{2187}
    + \frac{4160}{27}\zeta_3\right)n_f - \frac{2408}{729}n_f^2,
\nonumber\\
\gamma_{33}^{(3)} &= \frac{206734549}{65610} + \frac{560}{27}\zeta_3 - \left(\frac{3126367}{10935}
    + \frac{5120}{27}\zeta_3\right)n_f - \frac{14722}{3645}n_f^2,
\nonumber\\
\gamma_{44}^{(3)} &= \frac{144207743479}{41006250} + \frac{9424}{405}\zeta_3 - \left(\frac{428108447}{1366875}
    + \frac{5888}{27}\zeta_3\right)n_f - \frac{418594}{91125}n_f^2,
\nonumber\\
\gamma_{55}^{(3)} &= \frac{183119500163}{47840625} + \frac{3328}{135}\zeta_3 - \left(\frac{1073824028}{3189375}
    + \frac{2176}{9}\zeta_3\right)n_f - \frac{3209758}{637875}n_f^2.
\end{align}
Note that the index $n$ enumerates elements in the Gegenbauer basis so that $\gamma_{nn} = \gamma(n + 1)$.
We have checked that the $n_f^2$ contributions to the off-diagonal matrix agree with the result obtained in
Ref.~\cite{VanThurenhout:2022nmx}~\footnote{The evolution kernel for the leading $n_f$ contribution in all orders can be found
in~\cite{VanThurenhout:2022hgd}.}.

\section{Summary}\label{sect:summary}
The theoretical description of  hard exclusive processes in QCD requires the knowledge of scale dependence of non-forward matrix elements of
local/non-local operators. It is described by the corresponding anomalous dimension matrix or evolution kernel, which is completely
determined by the forward anomalous dimensions at $\ell$~loops and an additional quantity, the conformal anomaly calculated in
$(\ell-1)$-loop approximation~\cite{Mueller:1991gd}. This arises from the hidden conformal symmetry present in the evolution kernels of the
$\overline{\text{MS}}$ scheme in QCD. The corresponding generators, however, receive quantum corrections and differ from the canonical
ones. The conformal anomaly, introduced by M\"uller, describes a non-trivial modification of the generator of special conformal
transformations. For the (axial-)vector nonsinglet twist-two operators the conformal anomaly was calculated at one- and two-loop accuracy
in Refs.~\cite{Belitsky:1998gc} and \cite{Braun:2016qlg}, respectively, and the evolution kernel for (axial-)vector operators are known now
at the three-loop level~\cite{Braun:2017cih}. For the transversity operators, the evolution kernel has been known with two-loop accuracy~\cite{Belitsky:1998gc,Belitsky:1999gu,Mikhailov:2008my}.

In this paper we have calculated the two-loop conformal anomaly for the generator of special conformal transformations for the transversity
operator in QCD. Using this result  and the corresponding forward three-loop anomalous dimensions calculated in~\cite{Velizhanin:2012nm,Blumlein:2021enk} we
have reconstructed the evolution kernel for the operators in question in non-forward kinematics. In addition we have derived the explicit
expression for the three-loop anomalous dimension matrix for the local operators containing up to six covariant derivatives. Extensions to
a higher number of covariant derivatives are straight forward. In this  form, our result is applicable to the renormalization of meson wave
functions and could be useful for lattice calculations of their first few moments.

\section*{Acknowledgments}
\addcontentsline{toc}{section}{Acknowledgments}

We are grateful to V.N.~Velizhanin for helpful discussions.
This work has been supported by the DFG through the Research Unit FOR 2926,
{\it Next Generation pQCD for Hadron Structure: Preparing for the EIC},
project number 40824754, DFG grant MO~1801/4-2 and
the ERC Advanced Grant 101095857 {\it Conformal-EIC}.

\appendix
\section*{Appendices}
\addcontentsline{toc}{section}{Appendices} 

\renewcommand{\theequation}{\Alph{section}.\arabic{equation}}
\renewcommand{\thetable}{\Alph{table}}
\setcounter{section}{0} \setcounter{table}{0}


\section{Results for two-loop diagrams}\label{sect:diagrams}
\subsection{Evolution kernel \label{app:kernel}}

The contributions to the evolution kernel from the diagrams in Fig.~\ref{fig:two-loop} (a)--(p) (including symmetric diagrams with the
interchange of the quark and the antiquark) can be written in the following form:
\begin{align}
  [\mathbb{H} \mathcal{O}](z_1z_2) &=
-4 \int_0^1\!d\alpha\!\int_0^{\bar\alpha} d\beta \Big[\chi(\alpha,\beta) + \chi^{\mathbb{P}}(\alpha,\beta) \mathbb{P}_{12}\Big]
\Big[\mathcal{O}(z_{12}^{\alpha},z_{21}^\beta)+\mathcal{O}(z_{12}^{\beta},z_{21}^\alpha)\Big]
\notag\\
&\quad
 -4 \int_0^1\!du \, h (u) \Big[2 \mathcal{O}(z_1,z_2) - \mathcal{O}(z_{12}^u,z_2) - \mathcal{O}(z_1,z_{21}^u)\Big]\,,
\end{align}
where $\mathbb{P}_{12}$ is the permutation operator. For any function $f(z_1,z_2)$
\begin{align}
  \mathbb{P}_{12} f(z_1,z_2) = f(z_2,z_1), &&
 \Big( \mathbb{P}_{12} \mathcal{O}(z_{12}^\alpha,z_{21}^\beta) = \mathcal{O}(z_{21}^\alpha,z_{12}^\beta) \Big).
\end{align}
One obtains (only the non-vanishing contributions are listed):
\begin{align}
  h_{(a)}(u) &= C_F^2\frac{\bar u}{u}\left[\ln u+1\right],
\nonumber\\
  h_{(b)}(u) &=
 C_F \frac{\bar u}{u}
 \Big[ (2C_A-\beta_0 ) \ln \bar u + \frac83 C_A  - \frac53 \beta_0 \Big],
\nonumber\\
  h_{(c)}(u) &=
\Big[C_F^2 - \frac12 C_FC_A\Big] \frac{\bar u}{u} \Big[\ln^2\bar u - 3 \frac{u}{\bar u}\ln u  + 3 \ln \bar u -\ln u - 1\Big],
\nonumber\\
  h_{(d)}(u) &= \frac12 C_F C_A
\frac{\bar u}{u} \left[ \frac12 \left(1-\frac{u}{\bar u}\right)\ln^2 u +\ln \bar u -3\right],
\nonumber\\
  \hspace{-2mm}h_{(e+f)}(u) &=
   2\, C_F^2 \frac{\bar u}{u} \left[ 2\big( \Li_2(1)-\Li_2(\bar u)\big) - \ln^2\bar u + 2 \frac{u}{\bar u}\ln u\right]
\nonumber\\
    &\quad
    + C_F C_A  \frac{\bar u}{u} \biggl[2 \big(\Li_2(\bar u)- \Li_2(u)\big) + \frac12 \ln^2\bar u
     - \frac12 \ln^2 u - \frac{1+ u}{\bar u}\ln u-2 \biggr],
\nonumber\\
  h_{(g)}(u) &= -  C_F C_A \frac{\bar u}{u}
\biggl[
   \Li_2(\bar u)-\Li_2(1)+1+\frac{1}{4} \ln^2\bar u
+\ln\bar u -\frac{1+u}{2\bar u}  \ln u \left(\frac12 \ln u +1\right)\biggr],
\nonumber\\
 h_{(j)}(u) &=\Big[C_F^2 - \frac12 C_FC_A\Big] \,\ln u\,,
\nonumber\\
  h_{(o)}(u) &=2\Big[C_F^2 - \frac12 C_FC_A\Big]\frac{\bar u}{u}
\biggl[-2\Li_2(u)+\frac{u}{\bar u}\ln u\ln\bar u-\frac12 \ln^2 \bar u-\frac{u}{\bar u}\ln u\biggr],
\nonumber\\
 h_{(p)}(u) &=C_FC_A\frac{\bar u}u
 \biggl[\Li_2( u)+\frac1{\bar u}\ln u\ln\bar u-\frac14 \ln^2\bar u-\frac{u}{4\bar u}\ln^2u-\frac{u}{\bar u}\ln u\biggr],
\end{align}
and
\begin{align}
\chi_{(i)}(\alpha,\beta)&= 
\frac16C_F(C_A-\beta_0)\delta(\alpha)\delta(\beta),
\notag\\
\chi_{(j)}(\alpha,\beta)&=\Big[C_F^2 - \frac12 C_FC_A\Big]
\delta(\alpha)\delta(\beta),
\notag\\
\chi_{(m)}(\alpha,\beta)& = \Big[C_F^2 - \frac12 C_FC_A\Big], 
\notag\\
\chi_{(o)}(\alpha,\beta)& =
-2\Big[C_F^2 - \frac12 C_FC_A\Big]\left[
\frac1{\bar\alpha} \ln\alpha-\frac1{\alpha}\ln\bar\alpha -\frac{\bar\tau}{\tau}\ln\bar\tau
+\big[2+\zeta_2-3\zeta_3\big] \delta(\alpha)\delta(\beta)\right],
\notag\\
\chi_{(p)}(\alpha,\beta)& =  C_F C_A\left[\frac1\alpha\ln\bar\alpha  -\frac1{\bar\alpha}\ln\alpha
 + \big[\zeta_2-2\big]\delta(\alpha)\delta(\beta)\right].
\end{align}
The nonvanishing contributions to $\overline\chi(\alpha,\beta)$ originate from two diagrams only:
\begin{align}
\overline\chi_{(m)}(\alpha,\beta) &= 
\Big[C_F^2 - \frac12 C_FC_A\Big],
\notag\\
\overline\chi_{(o)}(\alpha,\beta) &=-2\Big[C_F^2 - \frac12 C_FC_A\Big]\, \bar \tau\ln\bar\tau\,.
\end{align}
We note here that the results for the $h$ functions are exactly the same as in the vector case~\cite{Braun:2016qlg}.

\subsection{Conformal anomaly\label{app:anomaly}}
Terms due to the conformal variation of the action can be written in the form
\begin{align}
\Delta S_+  &= \frac12 \mathbb{H}(z_1+z_2) + z_{12} \Delta_+ \,,
\end{align}
where $\mathbb{H}$ is the corresponding contribution to the evolution kernel.
The contributions to $\Delta_+$ from the diagrams in Fig.~\ref{fig:two-loop}
(including symmetric diagrams with the interchange of the quark and the antiquark) can be brought to the following form:
%
below we list the non-vanishing contributions only {\allowdisplaybreaks
  \begin{align}
\label{two-loop-anomaly-single}
  \varkappa_{(a)}(t) &= C_F^2\left[\frac1t+\frac{1+\bar t}{t}\ln t\right],
\nonumber\\
  \varkappa_{(b)}(t) &=  - 2 C_F \frac{\bar t}{t}
 \Big[ (\beta_0 -2 C_A ) \ln \bar t -  \frac83 C_A  + \frac53 \beta_0 \Big],
\nonumber\\
\varkappa_{(c)}(t) &= \Big[C_F^2 - \frac12 C_FC_A\Big]\Big[t \ln ^2 t  + \frac{2\bar t}{t}\ln ^2\bar t  + \frac{6\bar t}{t} \ln \bar t
- \frac{\bar t}{t} (3t+2) \ln t  - 9 t + 8 -\frac{1}{t}\Big],
\nonumber\\
\varkappa_{(d)}(t) &= C_F C_A\biggl\{ \frac{\bar t}{t} \left[ 
\frac{1-2t}{2\bar t}\ln^2 t +\ln \bar t -3\right]
\notag\\
&\quad
+\frac12\biggl[ \frac12\ln^2 t-\bar t\ln^2\bar t+\frac{t^2-\bar t}{t}\ln t-2\bar t\ln\bar t-1-\bar t\biggr]
\biggr\},
\nonumber\\
\varkappa_{(e+f)}(t) &=
 -4C_F^2\bigg\{ t\Big[\Li_2(t)-\Li_2(1)\Big] + 2 \frac{\bar t}{t} \Big[\Li_2(\bar t)-\Li_2(1)\Big]
 + \frac{\bar t}{t} \ln^2 \bar t + \frac12 t \ln^2 t + 2 \bar t \ln\bar t
\nonumber\\&\quad
 - \frac32(1-2t)\ln t + 2 \biggr\}
+ C_F C_A \frac{\bar t}{t} \biggl\{
4 \Big[\Li_2(\bar t)-\Li_2(t)\Big] + \frac12(2+t)\ln^2\bar t
\nonumber\\&\quad
- \left(1-\frac{t^2}{2\bar t}\right)\ln^2 t
- 2(1-2t)\ln\bar t - \left(5t+ \frac{1}{\bar t}\right)\ln t - 3 + 2t\biggr\},
\nonumber\\
\varkappa_{(g)}(t) &= C_F C_A \frac{\bar t}{t}\biggl\{ t\Big[\Li_2(\bar t) - \Li_2(1)\Big]+ \frac14 t \ln^2\bar t + \frac14(2\!+\!t)\ln^2 t -
           (3\!-\!t)\ln\bar t
\nonumber\\&\quad
+ \frac12 \left(1-\frac{t^2}{\bar t}\right)\ln t  -\bar t -\frac32\biggr\},
\nonumber\\
  \varkappa_{(j)}(t) &= \Big[C_F^2 - \frac12 C_FC_A\Big]\Big[-t\ln t -1\Big],
\nonumber\\
  \varkappa_{(o)}(t) &= \Big[C_F^2 - \frac12 C_FC_A\Big]
\Big\{
\frac4{\bar t}\Big[\Li_2(t)-\Li_2(1)\Big]-4t\Big[\Li_2(t)-\Li_2(1)\Big]+4\bar t\Li_2(1)
\notag\\&\quad
-2t\ln t\ln\bar t +\frac t{\bar t}\ln^2t+\bar t
\ln^2\bar t-4t\ln\bar t+\frac{2t}{\bar t}(2-3t)\ln t+2
\Big\},
\nonumber\\
  \varkappa_{(p)}(t) &= C_F C_A \Big\{
\frac{2t}{\bar t}\Big[\Li_2(t)-\Li_2(1)\Big]+\bar t \Big[\Li_2(\bar t)-\Li_2(1)\Big]
-t\ln t\ln\bar t
\nonumber\\&\quad
+\frac14\bar t\ln^2\bar t
+\frac14\frac{t(3-t)}{\bar t}\ln^2t
-\frac{t^2}{\bar t}\ln t+\frac12\ln t-\frac{1+t}{t}\ln\bar t + 1\Big\}.
\end{align}
Note here that all $\varkappa$ functions are exactly the same as in the vector case.
The function $\overline\omega(\alpha,\beta)$ receives contributions from two diagrams only:
\begin{align}
\label{two-loop-anomaly-double-p}
 \overline \omega_{(m)}(\alpha,\beta) &= -2 \left[C_F^2-\frac12C_F C_A\right]\,\beta,
\nonumber\\
\overline  \omega_{(o)}(\alpha,\beta) &=
  4 \left[C_F^2-\frac12C_F C_A\right]\,\Biggl\{
  \frac\alpha{\bar\alpha}\left[\Li_2\left({\frac\alpha{\bar\beta}}\right)-\Li_2(\alpha)\right]
  -\alpha\bar\tau\ln\bar\tau-\frac1{\bar\alpha}\ln\bar\alpha\ln\bar\beta-\frac{\beta}{\bar\beta}\ln\bar\alpha
\Biggr\}.
%
\end{align}
}
%
The non-vanishing contributions to $\omega(\alpha,\beta)$ are
\begin{align}
\label{two-loop-anomaly-double}
  \omega_{(m)}(\alpha,\beta) &= 2 \Big[C_F^2 - \frac12 C_FC_A\Big]\beta,
\nonumber\\
  \omega_{(o)}(\alpha,\beta) &= -4\Big[C_F^2 - \frac12 C_FC_A\Big]
 \biggl\{\frac{\bar\alpha}\alpha\Big[\Li_2(\beta/\bar\alpha)-\Li_2(\alpha)-\Li_2(\beta)\Big]
  -\frac\alpha{\bar\alpha}\Big[\Li_2(\alpha)-\zeta_2\Big]
\nonumber\\&\quad
- \frac14\frac{\bar\alpha}\alpha\ln^2\bar\alpha-\frac14 \frac\alpha{\bar\alpha}\ln^2\alpha+\ln\alpha\ln\bar\alpha
+ \alpha\frac{\bar\tau}{\tau}\ln\bar\tau
+\frac\alpha{\bar\alpha}\ln\alpha
-\frac12\frac{\bar\beta}{\bar\alpha}\ln\alpha -\frac12\frac\beta\alpha\ln\bar\alpha
\biggr\},
\nonumber\\
  \omega_{(p)}(\alpha,\beta) &= 
C_FC_A
\biggl\{
\frac{2}{\bar\alpha} \Big[\Li_2(\alpha)-\Li_2(1)\Big]+\frac{2}{\alpha} \Big[\Li_2(\bar \alpha)-\Li_2(1)\Big]
+\frac12\frac{\bar\alpha}\alpha\ln^2\bar\alpha
+\frac12\frac\alpha{\bar\alpha}\ln^2\alpha
\notag\\
&\quad
-\frac{2}{\bar\alpha}\ln\alpha - \frac{2}{\alpha}\ln\bar\alpha
+\frac\beta\alpha\ln\bar\alpha +\frac{\bar\beta}{\bar\alpha}\ln\alpha
\biggr\}.
%
%
%
\end{align}

\section{ $\mathrm X$ kernel}\label{sect:X}
In this appendix we present the results for the two-loop kernel $\mathrm X^{(2)}$ (the one-loop result is given in Eq.~\eqref{X1}).
The kernel $\mathrm X^{(2)}$ is defined as the solution of the Eq.~\eqref{Xeq-2}. For the technical use this relation can be seen as a
differential equation for the integration kernel.
In general, for an arbitrary integral operator $\mathrm{F}$ of the form
\begin{align}
    \left[\mathrm{F}f\right](z_1, z_2) & = F_{\text{const}}f(z_1, z_2)
     + \int_0^{1}d\alpha\int_{0}^{\bar{\alpha}}d\beta\,h(\alpha, \beta) f(z_{12}^{\alpha}, z_{21}^{\beta})
     \nonumber\\
     &\quad + \int_0^{1}d\alpha\dfrac{\bar{\alpha}}{\alpha}h^\delta(\alpha)\left(2f(z_1, z_2)
      - f(z_{12}^{\alpha},z_2) - f(z_1, z_{21}^{\alpha})\right),
\end{align}
its commutator with the generator $S_+^{(0)}$ has the form
\begin{align}
    \left[S_{+}^{(0)}, \mathrm{F}\right]f & = z_{12}\int_{0}^{1}d\alpha\int_{0}^{\bar{\alpha}}d\beta\,
    \left(\alpha\bar{\alpha}\partial_\alpha - \beta\bar{\beta}\partial_\beta\right)h(\alpha, \beta)f(z_{12}^\alpha, z_{21}^{\beta})
     \nonumber \\
     &\quad - z_{12}\int_0^1 d\alpha\,\bar{\alpha}^2\partial_\alpha h^\delta(\alpha)\left(f(z_{12}^\alpha, z_2) - f(z_1, z_{21}^\alpha)\right).
\end{align}
The kernel $\mathrm{X}^{(2)}$  can be written as a sum of three
terms corresponding to the three contributions on the right hand side of Eq.~\eqref{Xeq-2}
\begin{equation}
    \mathrm{X}^{(2)} = \mathrm{X}^{(2)}_{\text{I}}
    + \mathrm{X}^{(2,1)}\left(\beta_0 + \dfrac{1}{2}\mathbf {H}_\text{inv}^{(1)}\right) - \frac{1}{2}\mathrm{X}^{(2,2)}.
\end{equation}
It is easy to see that the operators $\mathrm{X}^{(2,1)}$ and $\mathrm{X}^{(2,2)}$ are exactly the same as in the vector
case~\cite{Braun:2017cih,Strohmaier:2018tjo}
\begin{equation}
    \mathrm{X}^{(2,1)}f(z_1,z_2) =
     -2C_F\int_0^1 d\alpha\,\left[\frac{\bar{\alpha}}{\alpha}\ln\bar{\alpha}
      + \ln\alpha\right]\Big[2f(z_1, z_2) - f(z_{12}^\alpha, z_2) - f(z_1, z_{21}^\alpha)\Big],
\end{equation}
and
\begin{align}
    &\mathrm{X}^{(2,2)}f(z_1, z_2) =
    \notag\\
    &=4C_F^2\Biggl\{ \int_0^1  d\alpha\int_0^1 du
    \left[\frac{\ln\bar{\alpha}}{\alpha}\left(\dfrac{1}{2}\ln\bar{\alpha} + 2\right)
       + \frac{\bar{u}}{u}\frac{\vartheta(\alpha)}{\bar{\alpha}}\right]
      \Big[2f(z_1, z_2) - f(z_{12}^{\alpha u}, z_2)
       - f(z_1, z_{21}^{\alpha u})\Big]
       \nonumber\\
       & + \int_0^1 d\alpha \int_0^{\bar{\alpha}}d\beta \,
        \Biggl[\frac{\vartheta_+(\alpha) + \vartheta_+(\beta)}\tau\left(f(z_{12}^\alpha, z_{21}^\beta)
        - f(z_1, z_{21}^\beta) - f(z_{12}^\alpha, z_2) + f(z_1, z_2)\right)
        \notag\\
        & +\Big(\vartheta_0(\alpha) + \vartheta_0(\beta)\Big) f(z_{12}^\alpha, z_{21}^\beta)\Biggr] \Biggr\},
\end{align}
where
\begin{align}
    \vartheta_+(\alpha) &=  -\frac{1}{\bar{\alpha}}\Big[\ln\alpha \ln\bar{\alpha}
    + 2\alpha\ln\alpha + 2\bar{\alpha}\ln\bar{\alpha}\Big],
\notag\\
    \vartheta_0(\alpha) &=  2\Big[\Li_3(\bar{\alpha}) - \Li_3(\alpha) - \ln\bar{\alpha}\Li_2(\bar{\alpha})
    + \ln\alpha\Li_2(\alpha)\Big] + \dfrac{1}{\alpha}\ln\alpha\ln\bar{\alpha} + \dfrac{2}{\alpha}\ln\bar{\alpha},
\\ \notag
    \vartheta(\alpha) &= \frac{\alpha}{\bar{\alpha}}\Big[\Li_2(\bar{\alpha}) - \ln^2\alpha\Big]
    - \frac{\bar{\alpha}}{2\alpha}\ln^2\bar{\alpha}
    + \left[\alpha - \dfrac{2}{\alpha}\right]\ln\alpha\ln\bar{\alpha}
     - \left[3 + \dfrac{1}{\bar{\alpha}}\right]\ln\alpha - (\alpha - \bar{\alpha})\dfrac{\bar{\alpha}}{\alpha} - 2.
\end{align}
The operator $\mathrm X_\text{I}$ obeys the following equation
\begin{align}
\label{x-1-equation}
    \left[S_{+}^{(0)}, \mathrm{X}_{\text{I}}^{(2)}\right] &= z_{12}\Delta_{+}^{(2)} + \dfrac{1}{4}\left[\mathbb{H}^{(2)},z_1 + z_2\right]
  \notag  \\
&    = z_{12}\Delta_+^{(2)} + \dfrac{1}{4}\left[\frac12 \mathrm{T}^{(1)}\mathbf H_\text{inv}^{(1)}+\left[\mathbf H_\text{inv}^{(1)},
 \mathrm{X}^{(1)}\right], z_1 + z_2\right]
     +\frac14 \left[\mathbf H_\text{inv}^{(2)}+\beta_0 \mathrm T_1^{(1)},z_1+z_2\right]
\end{align}.
The solution can be written as
\begin{align}
\mathrm{X}_{\text{I}}^{(2)} &=  \mathrm{X}_{\text{IAB}}^{(2)} 
+ \frac{1}{4}\left(\mathrm{T}^{(2)}_1
     + \frac12 \beta_0\mathrm {T}^{(1)}_{2}\right)\,.
\end{align}
The last term in this equation corresponds to the third term in Eq.~\eqref{x-1-equation}\,\footnote{We note here that our definition of the
operators $\mathrm T_n^{(k)}$ differs from that in Ref.~\cite{Braun:2017cih}}. We also note that the combination $\mathbf
H_\text{ninv}=\frac12 \mathrm{T}^{(1)}\mathbf{H}_\text{inv}^{(1)}+\left[\mathbf{H}_\text{inv}^{(1)}, \mathrm{X}^{(1)}\right]$ corresponds
to the non-invariant $C_F^2 $ part of the two-loop kernel and has the form,
\begin{align}
\mathbf H_\text{ninv} f(z_1,z_2) &= 8 C_F^2\Biggl( \int_0^1 d\alpha
 \frac{\bar\alpha}\alpha\ln\bar\alpha\left(\frac32-\ln\bar\alpha + \frac{1+\bar\alpha}{\bar\alpha}\ln\alpha\right)
\Big(f(z_{12}^\alpha,z_2)+ f(z_1,z_{21}^\alpha)\Big)
\notag\\
&\quad +\int_0^1d\alpha\int_0^{\bar\alpha} d\beta \left(\frac1{\bar\alpha}\ln\alpha-\frac1\alpha\ln\bar\alpha
+(\alpha\leftrightarrow \beta)\right)
f(z_{12}^\alpha,z_{21}^\beta)\Biggr)\,.
\end{align}
Since the two-loop anomaly $\Delta_+^{(2)}$ is also known one can easily find $\mathrm{X}_{\text{IAB}}^{(2)}$, which is convenient to
represent as a sum of two terms
\begin{align}
\mathrm{X}_{\text{IAB}}^{(2)}=\mathrm{X}_{\text{IA}}^{(2)}+\mathrm{X}_{\text{IB}}^{(2)}.
\end{align}
The first term $\mathrm{X}_{\text{IA}}^{(2)}$ contains all  contributions where at least one argument of the function remains intact.
Moreover, this term is exactly the same as in the vector case,
\begin{align}
    \mathrm{X}_{\text{IA}}f(z_1, z_2) & =
    \int_0^1 du \frac{\bar{u}}{u}\int_0^1\frac{d\alpha}{\bar{\alpha}}\Big(\varkappa(\alpha)
     - \varkappa(1)\Big)\Big(2f(z_1,z_2) - f(z_{12}^{\alpha u}, z_2) - f(z_1, z_{21}^{\alpha u})\Big) +
     \nonumber \\
     &\quad + \int_0^1 d\alpha \,\xi_{\text{IA}}(\alpha) \Big(2f(z_1, z_2) - f(z_{12}^\alpha,z_2) - f(z_1, z_{21}^\alpha)\Big),
\end{align}
where $\varkappa(\alpha)$  can be found in~\eqref{varkappa} and
\begin{align}
    \xi_{\text{IA}}(\alpha) = &2C_F^2\frac{\bar{\alpha}}{\alpha}\left(-\Li_3(\bar{\alpha})
    + \ln\bar{\alpha}\Li_2(\bar{\alpha}) + \frac{1}{3}\ln^3\bar{\alpha} + \Li_2(\alpha)
     + \frac{1}{\bar{a}}\ln\alpha\ln\bar{\alpha} - \frac{1}{4}\ln^2\bar{\alpha}\right.
     \nonumber \\ &\left.-\frac{3\alpha}{\bar{\alpha}}\ln\alpha - 3\ln\bar{\alpha}\right)
     + {\frac{C_F}{N_c}}\left(\ln\alpha + \frac{\bar{\alpha}}{\alpha}\ln\bar{\alpha}\right).
\end{align}
The result for the second term $\mathrm X^{(2)}_\text{IB}$ reads
\begin{align}
    \mathrm{X}_{\text{IB}}^{(2)} &= \int_0^1 d\alpha \int_0^{\bar{\alpha}}d\beta \left(C_F^2\,\xi_{P}(\alpha,\beta)
    +\frac{C_F}{N_c}\left(\xi_{NP}(\alpha,\beta)
     + {\overline \xi_{NP}}(\alpha,\beta)\mathbb{P}_{12}\right)\right)f(z_{12}^\alpha, z_{21}^\beta),
\end{align}
where
\begin{align}
 \xi_P(\alpha, \beta)&=4\left(-\Li_3(\alpha) + \ln\alpha\Li_2(\alpha) + \frac{1}{\bar{\alpha}}\left(\Li_2(\alpha) - \zeta_2
     + \frac{1}{4}\ln^2\alpha - \ln\alpha\right) +(\alpha\leftrightarrow\beta)\right)
     \notag\\
     &\quad  -(\alpha,\beta \leftrightarrow \bar\alpha,\bar\beta),
\notag\\
\xi_{NP}(\alpha, \beta)  & = -\frac{2}{\alpha}\left(\Li_2\left(\frac{\beta}{\bar{\alpha}}\right) - \Li_2(\beta)
    - \Li_2(\alpha) + \Li_2(\bar{\alpha}) - \zeta_2\right) - \ln\bar{\alpha}
    + (\alpha \leftrightarrow \beta),
\notag\\
\overline\xi_{NP}(\alpha,\beta) &=
     \frac{2}{\bar{\alpha}}\left(\Li_2\left(\frac{\alpha}{\bar{\beta}}\right) -\Li_2(\alpha) - \ln\bar{\alpha}\ln\bar{\beta}\right)
     - \ln\bar{\alpha} + (\alpha \leftrightarrow \beta).
\end{align}
Note that the integral kernel $\xi_P(\alpha, \beta)$ corresponds to $z_{12}\Delta_+^{(2)}$ and the second term in
Eq.~\eqref{x-1-equation} while $\xi_{NP}(\alpha, \beta)$ and $\overline{\xi}_{NP}$ correspond only to $z_{12}\Delta_{+}^{(2)}$.

\section{Parity invariant harmonic sums and integration kernels}\label{sect:Omegas}

In this appendix we give explicit expression for the harmonic sums which appears in the three-loop invariant
kernel in Eq.~\eqref{three-loop-int-kernel}.
The sums can be  divided in two groups with the respect to their signature, $\Pi_i^k \sign(m_i) = \pm 1$,
\begin{align}\label{Omega+}
    \Omega_3 &= S_3 - \zeta_3,
\nonumber\\
   \Omega_5 &=S_5 - \zeta_5,
\nonumber\\
    \Omega_{3,1} &= S_{3,1} - \frac{1}{2}S_{4} - \dfrac{3}{10}\zeta^2_2,
\nonumber\\
    \Omega_{1,3}&= S_{1,3} - \frac{1}{2}S_{4} + \dfrac{3}{10}\zeta_2^2 - \zeta_3S_1,
\nonumber\\
    \Omega_{-2,-2}&=S_{-2,-2} - \dfrac{1}{2}S_{4} + \frac{\zeta_2}{2}S_{-2} - \dfrac{\zeta_2^2}{8},
\nonumber\\
    \Omega_{1,3,1} &= S_{1,3,1} - \frac{1}{2}S_{4,1} - \frac{1}{2}S_{1,4}
    +  \frac{1}{4}S_{5} - \frac{3}{10}\zeta_2^2S_1 + \dfrac{3}{4}\zeta_5,
\nonumber\\
    \Omega_{1,1,3}&= S_{1,1,3} - \frac{1}{2}S_{2,3} - \frac{1}{2}S_{1,4} +  \frac{1}{4}S_{5}
    - \frac{\zeta_5}{2} + \frac{3}{10}\zeta_2^2S_1 + \frac{\zeta_3}{2}S_2 - \zeta_3S_{1,1},
\nonumber\\
    \Omega_{-2,-2, 1} &=S_{-2,-2,1} - \frac{1}{2}S_{-2,-3} - \frac{1}{2}S_{4,1} + \frac{1}{4}S_5 + \frac{1}{4}\zeta_3S_{-2} + \frac{1}{16}\zeta_5,
\nonumber\\
    \Omega_{-2, 1, -2} &= S_{-2,1,-2} - \frac{1}{2}S_{-2,-3} - \frac{1}{2}S_{-3,-2} + \frac{1}{4}S_5
    \nonumber\\ &
    - \frac{\zeta_2}{4}S_{-3} + \frac{1}{2}\zeta_2 S_{-2,1} - \frac{1}{4}\zeta_3S_{-2} + \frac{1}{8}\zeta_2\zeta_3 -\frac{3}{8}\zeta_5,
\nonumber\\
    \Omega_{1,-2, -2} &= S_{1,-2,-2} - \frac{1}{2}S_{-3,-2} - \frac{1}{2}S_{1,4} + \frac{1}{4}S_5
     \nonumber \\ &-\frac{\zeta_2}{4}S_{-3} + \frac{\zeta_2}{2} S_{1,-2} + \frac{1}{8}\zeta_2^2 S_1 - \frac{1}{8}\zeta_2\zeta_3
     + \frac{1}{16}\zeta_5,
\end{align}
and
\begin{align}\label{Omega-}
    {\Omega}_{-2} & = (-1)^N\left(S_{-2} + \dfrac{\zeta_2}{2}\right)
\nonumber \\
    {\Omega}_{-4} & =(-1)^N\left(S_{-4} + \dfrac{7\zeta^2_2}{20}\right)
\nonumber \\
    {\Omega}_{1,-2} & =(-1)^N\left(S_{1,-2} -\frac{1}{2}S_{-3} - \dfrac{\zeta_3}{4} + \dfrac{\zeta_2}{2}S_1\right)
\nonumber \\
    {\Omega}_{-2,1} & = (-1)^N\left(S_{-2,1} - \frac{1}{2}S_{-3} + \dfrac{\zeta_3}{4}\right)
\nonumber \\
    {\Omega}_{1,-4} & = (-1)^N\left(S_{1,-4} - \dfrac{1}{2}S_{-5}+ \dfrac{7}{20}\zeta_2^2S_1 - \dfrac{11}{8}\zeta_5
     + \dfrac{1}{2}\zeta_2\zeta_3\right)
\nonumber \\
    {\Omega}_{-4,1} & = (-1)^N\left(S_{-4,1} - \dfrac{1}{2}S_{-5} - \dfrac{1}{2}\zeta_2\zeta_3 + \dfrac{11}{8}\zeta_5\right)
\nonumber \\
    {\Omega}_{3,-2} & = (-1)^N\left(S_{3,-2} - \dfrac{1}{2}S_{-5} + \dfrac{1}{2}\zeta_2S_3 + \dfrac{9}{8}\zeta_5
    - \dfrac{3}{4}\zeta_2\zeta_3\right)
\nonumber \\
    {\Omega}_{1,-2,1} & = (-1)^N\left(S_{1,-2,1} -\frac{1}{2}S_{-3,1} -\frac{1}{2}S_{1,-3} + \frac{1}{4}S_{-4} + \dfrac{\zeta_3}{4}S_1
     - \dfrac{\zeta_2^2}{80}\right)
\nonumber \\
   {\Omega}_{1,1,-2,1} & = (-1)^N\Big(S_{1,1,-2,1} - \frac{1}{2}S_{1,-3,1} -  \frac{1}{2}S_{1,1, -3} - \frac{1}{2}S_{2,-2,1}
    + \frac{1}{4}S_{-4,1} + \frac{1}{4}S_{-4,1}+ \frac{1}{4}S_{2,-3}
     \nonumber\\
    &\quad
     + \frac{1}{4}S_{1,-4} - \frac{1}{8}S_{-5}
    +  \dfrac{\zeta_3}{4}S_{1,1} - \dfrac{\zeta_2^2}{80}S_1 - \dfrac{\zeta_3}{8}S_{2} + \dfrac{1}{8}\zeta_5 - \dfrac{1}{16}\zeta_2\zeta_3\Big).
\end{align}
Each sum $\Omega_{\vec{m}}$ is associated with the integral kernel $h_{\vec{m}}$ as follows
\begin{align}
    \int_0^{1}d\alpha\int_0^{\bar\alpha} d\beta\, h_{\vec{m}}(\tau)(1 - \alpha - \beta)^{N - 1} = \Omega_{\vec{m}}(N).
\end{align}
Below we list the  integral kernels corresponding to the sums~\eqref{Omega+} and \eqref{Omega-}
\begin{align}
    h_{3} & = -\dfrac{1}{2}\dfrac{\bar{\tau}}{\tau}\mathrm{H}_1 & h_{-2} &= \dfrac{1}{2}\bar{\tau}
\nonumber \\
    h_{5} & = -\dfrac{1}{2}\dfrac{\bar{\tau}}{\tau}\left(\mathrm{H}_{111} + \mathrm{H}_{12}\right)
    &
    h_{-4} &= \dfrac{1}{2}\bar{\tau}\left(\mathrm{H}_{11} + \mathrm{H}_{2}\right)
\nonumber \\
    h_{13} & = \dfrac{1}{4}\dfrac{\bar{\tau}}{\tau}\left(\mathrm{H}_2 + \mathrm{H}_{11}\right)
    &
     h_{1,-2} & = -\dfrac{\bar{\tau}}{4}\mathrm{H}_1
\nonumber \\
    h_{31} & = \dfrac{1}{4}\dfrac{\bar{\tau}}{\tau}\left(\mathrm{H}_{11} + \mathrm{H}_{10}\right)
    &
    h_{-2,1}& = -\dfrac{\bar{\tau}}{4}\left(\mathrm{H}_1 + \mathrm{H}_0\right)
\nonumber \\
    h_{113} & = -\dfrac{1}{8}\dfrac{\bar{\tau}}{\tau}\left(\mathrm{H}_{21} + \mathrm{H}_{111} + \mathrm{H}_{12} + \mathrm{H}_3\right)
    &
    h_{3,-2} & = -\dfrac{1}{4}\bar{\tau}\left(\mathrm{H}_{21} + \mathrm{H}_{111}\right)
\nonumber \\
    h_{131} & = -\dfrac{1}{8}\dfrac{\bar{\tau}}{\tau}\left(\mathrm{H}_{20} + \mathrm{H}_{110} + \mathrm{H}_{21} + \mathrm{H}_{111}\right)
    &
    h_{-4,1} & = -\dfrac{1}{4}\bar{\tau}\left(\mathrm{H}_{21} + \mathrm{H}_{20} + \mathrm{H}_{111} + \mathrm{H}_{110}\right)
\nonumber \\
    h_{-2,-2} & = \dfrac{1}{4}\dfrac{\bar{\tau}}{\tau}\mathrm{H}_{1,1}
    &
    h_{1, -4} &= \dfrac{1}{4}\bar{\tau}\left(\mathrm{H}_{111} - \mathrm{H}_{101}\right)
\nonumber \\
   h_{-2,-2,1} & = -\dfrac{1}{8}\dfrac{\bar{\tau}}{\tau}\left(\mathrm{H}_{111} + \mathrm{H}_{110}\right)
   &
   h_{1, -2, 1} & = \dfrac{1}{8}\bar{\tau}\left(\mathrm{H}_{11} + \mathrm{H}_{10}\right)
\nonumber \\
    h_{-2,1,-2} & = \dfrac{1}{8}\dfrac{\bar{\tau}}{\tau}\mathrm{H}_{111}
    &
    h_{1, 1, -2, 1} &= -\dfrac{1}{16}\bar{\tau}\left(\mathrm{H}_{111} + \mathrm{H}_{110}\right),
\nonumber \\
    h_{1,-2,-2} & = -\dfrac{1}{8}\dfrac{\bar{\tau}}{\tau}\left(\mathrm{H}_{111} + \mathrm{H}_{21}\right),
\end{align}
where $\mathrm{H}_{\vec{m}} = \mathrm{H}_{\vec{m}}(\tau)$ are HPLs.


\providecommand{\href}[2]{#2}\begingroup\raggedright\endgroup

\end{document}